\documentclass[11pt]{article}

\usepackage{graphicx}              
\usepackage{amsmath}               
\usepackage{amsfonts}              
\usepackage{amsthm}                
\usepackage{color}

\makeatletter
\@addtoreset{equation}{section}
\makeatother

\newcommand{\be}{\begin{eqnarray}}
\newcommand{\ee}{\end{eqnarray}}
\newcommand{\nn}{\nonumber}
\newcommand{\diag}{\mbox{diag}}

\newcommand{\equ}{\stackrel{!}{=}}
\newcommand{\sr}{{\sigma_r}}
\newcommand{\st}{{\sigma_\theta}}
\newcommand{\sph}{{\sigma_\phi}}
\newcommand{\PL}{{P_\Lambda}}
\newcommand{\PR}{{P_R}}

\begin{document}

\title{LTB spacetimes in terms of Dirac observables}
\author{
{\sf Kristina Giesel$^{3}$}\thanks{{\sf giesel(at)nordita(dot)org}},
{\sf Johannes Tambornino$^{1}$}\thanks{{\sf 
johannes(dot)tambornino(at)aei(dot)mpg(dot)de}},
{\sf Thomas Thiemann$^{1,2}$}\thanks{{\sf 
thiemann(at)aei(dot)mpg(dot)de, 
tthiemann(at)perimeterinstitute(dot)ca}}\\
\\
{\sf $^1$ MPI f. Gravitationsphysik, Albert-Einstein-Institut,} \\
           {\sf Am M\"uhlenberg 1, 14476 Potsdam, Germany}\\
\\
{\sf $^2$ Perimeter Institute for Theoretical Physics,} \\
{\sf 31 Caroline Street N, Waterloo, ON N2L 2Y5, Canada}\\
\\
{\sf $^3$ Nordic Institute for Theoretical Physics (NORDITA),} \\
{\sf Roslagstullsbacken 23, 106 91, Stockholm, Sweden}
}
\date{{\tiny Preprint AEI-2009-053}}

\maketitle
\thispagestyle{empty}

\begin{abstract} The construction of Dirac observables, that 
is gauge invariant objects, in General Relativity is technically more 
complicated than in other gauge theories such as the 
standard model due to its more complicated gauge group which  
is closely related to the group of spacetime diffeomorphisms.
 However, the explicit and usually cumbersome expression of Dirac 
observables in terms of gauge non invariant quantities is irrelevant
if their Poisson algebra is sufficiently simple. Precisely that can be 
achieved by employing the relational 
formalism and a specific type of matter proposed originally
by Brown and Kucha{\v r}, namely pressureless dust 
fields. Moreover one is able to derive a compact expression for 
a physical Hamiltonian that drives 
their physical time evolution. The resulting gauge invariant Hamiltonian 
system 
is obtained by Higgs -- ing the dust scalar fields and has an infinite 
number of conserved charges which force the Goldstone bosons to 
decouple from the evolution.  
In previous publications we have shown that explicitly for cosmological
perturbations. In this 
article we analyse the spherically symmetric sector of the theory  
and it turns out that the solutions are in 
one--to--one correspondence with the class of Lemaitre--Tolman--Bondi 
metrics.  
Therefore the theory is capable of properly describing the whole 
class of gravitational experiments that rely on the assumption of  
spherical symmetry. 
\end{abstract}

\newpage
\section{Introduction}
Until today General Relativity (GR) is the most successful theory in 
describing  observations involving gravitational interactions on 
macroscopic scales.  It has been tested in various experimental settings 
and so far  many predictions of GR have been confirmed to great 
accuracy. But next to the complicated highly non -- linear  structure of 
Einstein's equations there is one aspect in GR which has to be treated 
with special care as compared to other field theories such as 
Maxwell's theory: the issue of observables.  Observables are those 
quantities that respect the gauge symmetry of the theory, for 
Electrodynamics this would simply be all $U(1)$ -- invariant quantities.  
In contrast for GR the gauge group  is closely related to the group of 
space time 
diffeomorphisms, denoted by  Diff($  M$), reflecting the independence of 
physics on space time coordinates.  We are dealing with a background 
independent theory where ``space'' and ``time'' do not have any a priori 
physical meaning.
\\
In the canonical picture of GR \cite{adm}, where one performs a (3+1) -- 
split of the four dimensional space time $(M, g)$, gauge invariance 
carries over to the condition that phase space functions have to commute 
with the Hamiltonian and spatial diffeomorphism constraints of 
GR\footnote{See 
\cite{dirac} or \cite{hentei} for a detailed description how to deal 
with gauge symmetries in physical theories.}. The reason why it is so 
complicated to extract the gauge--invariant content of GR is that the 
Poisson algebra formed by the constraints of GR, the so called Dirac 
algebra, is an extremely difficult one. It is not only non--Abelian but 
is not even an honest Lie--algebra: instead of structure constants it 
involves structure functions, i.e.,  phase space dependent quantities.
However, even if observables were known there would still be a 
conceptual task to solve which is often referred to as the problem of 
time in GR \cite{ProbTime1,ProbTime2} and concerns the evolution of 
observables. Speaking about dynamics of GR we usually refer to 
Einstein's equations. In the canonical picture these can be obtained 
from a canonical Hamiltonian and the Hamiltonian and diffeomorphism 
constraint equations. However, as a consequence of diffeomorphism 
invariance the canonical Hamiltonian is a linear combination of 
constraints only\footnote{In the absence of boundaries such as 
in asymptotically flat situations.}. 
Hence, the Hamiltonian equations generated by this canonical Hamiltonian 
will be identical to zero in the case of observables. 
Consequently, Einstein's equations do not describe what we usually would 
call physical evolution but rather describe how the metric changes under 
gauge transformations. What is needed for observables is a gauge 
invariant version of Einstein's equation, generated by a so called 
physical Hamiltonian and describing non -- trivial evolution of 
observables.
\\
A framework that allows to construct  observables and analyse their 
evolution for constrained systems such as GR is the so called Relational 
Formalism. The basic idea is to take the background independent nature 
of GR seriously and define observables not with respect to unphysical 
space time points in ${  M}$ but to use relations between dynamical 
fields instead. These ideas date back to the seminal work of Bergmann 
and Komar 
\cite{bergmann, komar, bergmann_komar} from the 1960s. Its conceptual 
foundations were very much improved in the 1990s, see  
\cite{rovelli1, rovelli2} and references therein. The 
corresponding mathematical framework was also developed in the 1990's,
see e.g. \cite{GaugeUnfixing} and references therein and rediscovered 
more recently in \cite{dittrich1, dittrich2}.  
Once appropriate dynamical fields are 
chosen as clocks\footnote{Whenever we talk of clocks we mean devices 
(fields) to measure temporal {\it and} spatial distances.} in order to 
give space and time a physical meaning one can at least formally write 
down an expression for observables associated to any phase space 
function $f$.  In \cite{dittrich_tambo1, dittrich_tambo2} a perturbative 
scheme to compute these observables was developed and its application to 
perturbation theory around Minkowski space and cosmological perturbation 
theory for different choices of clocks was discussed.
However, the explicit form is a power series in these clock variables 
with coefficients involving multiple Poisson brackets of the constraints 
and $f$ generally leading to a rather complicated dependence on the 
physical time parameter that in most cases cannot be written down in a 
closed analytic form.  

However, as pointed out for instance in \cite{thiemann1}, the precise 
expression of the Dirac observables in terms of non gauge inavriant 
objects, which generically involves a hopelessly complicated, infinite 
series of multiple Poisson brackets, is in fact irrelevant from the 
point of view of the physical (or reduced) phase space that we are 
actually interested in. What is relevant is the Poisson algebra of those 
Dirac observables and the physical Hamiltonian which drives their 
physical time evolution. General expressions for an arbitary first class
system and for general choices of clocks were derived in 
\cite{thiemann1}. From a practical point of view and especially with 
regard to quantisation it is of 
course desirable to 
have a sufficiently simple gauge invariant Poisson algebra 
$\mathfrak{A}_{{\rm phys}}$ and a compact 
expression for the resulting physical Hamiltonian ${\rm H}_{{\rm 
phys}}$. The structure of both $\mathfrak{A}_{{\rm phys}},\;\;{\rm 
H}_{{\rm phys}}$ depends sensitively on the choice of clocks and for 
generic choices both are beyond mathematical control. One
therefore should analyse first which type of matter simplifies 
their structure.     

First steps in this direction have been 
performed in \cite{thiemann2} where a phantom scalar field, i.e. a 
scalar field with negative kinetic energy term, has been used as a clock 
and the corresponding physical Hamiltonian has been derived. 
The choice of a scalar field with vanishing potential as a clock field 
was motivated entirely
by mathematical considerations based on the Brown -- Kucha{\v r} 
mechanism \cite{brown_kuchar} which allows to deparametrise the 
Hamiltonian constraint. From a physical point of view,   
introducing some ad hoc matter component that adopts the role of 
clocks for GR might look artificial and even dangerous, especially if 
that matter is in 
conflict with the ususal energy conditions that stabilise the system.
However, the way that the relational formalism works is in fact very 
similar to the Higgs mechanism: In the gauge invariant formulation, 
the scalar field completely disappears. What remains is the 
corresponding physical Goldstone boson and what one has to worry about 
is that this additional degree of freedom is not in conflict with 
observation, in particular, that the energy conditions hold in the 
gauge invariant description. 

A more physically motivated choice of clocks which have the additional
advantage of implementing the Brown -- Kucha\v{r} mechanism is defined
by the original Lagrangian for pressure free dust due to Brown and 
Kucha\v{r} \cite{brown_kuchar}. This choice of matter is a sense 
physically distinguished because it can be considered in a precise 
sense as a congruence of mutually non interacting, freely falling 
observers 
which only interact gravitationally and which define the 
dynamical reference frame for GR.
From the mathematical point of view, the essential point in the 
construction is that the 
constraints of the coupled system, including gravity, matter and dust, 
can be written in a deparametrised form\footnote{A system of constraints 
$C_I$ is said to be deparametrisable if one can find a local coordinate 
chart on phase space with two mutually commuting sets of canonical pairs 
denoted by $(q^a, p_a)$ and $ (T^I, P_I)$ such that in this chart the 
constraints  can be written in the locally equivalent form $C_I = P_I + 
h_I$ where $h_I$ depends only on $(q^a, p_a)$. Equivalence has to be 
understood in the sense that the constraint hypersurface, spanned by the 
deparametrised constraints is exactly the same as the one spanned by the 
original set of constraints.}. 
As was shown explicitly in \cite{ghtw1}, this achieves the goal of 
drastically 
simplifying the Poisson
algebra of physical observables and the physical Hamiltonian 
and furthermore ensures that the associated 
physical Hamiltonian is time independent. In contrast to 
\cite{brown_kuchar}, the requirement that the physical Hamiltonian should 
be positive definite, or equivalently physical time is required to run 
forward rather than backwards, demands the dust needs to be phantom 
dust. This is exactly the same reason why the phantom occurred in 
the previous work \cite{thiemann2} and ensures that the energy 
momentum tensor of observable matter obeys the usual energy conditions. 
It was also shown that the resulting
theory is in good agreement with current cosmological 
observations in 
\cite{ghtw2} as far as FRW space times as well as perturbations around 
FRW are considered. The reason for why that happens is the afore 
mentioned Higgs
mechanism by which the phantom dust completely disappears from the 
physical particle spectrum together with another nice feature of this
particular choice of clocks which is absent for generic choices:
The physical Hamiltonian system posesses an infinite number of conserved 
charges which can be considered as physical energy and momentum 
densities of the dust respectively and which enforce the corresponding 
gravitational Goldstone 
boson modes to decouple in a mathematically precise sense from the 
physical
time evolution. In particular, it is consistent to tune those charges 
to be arbitarily small so that their corrections to Einstein's equations 
for the non Goldstone modes are also arbitrarily small. In this way the 
dust comes as close as 
possible to the mathematical idealisation of a test observer while 
taking its gravitational interaction into account. Of 
course, one could also use gravitational clocks, that is certain 
components of the metric tensor and thus avoid the Goldstone modes 
altogether. However, while that works well for the linearised 
theory, taking into account the full non linearities of Einstein's 
theory leads to equations of motion for the 
physical graviational modes that have no resemblance with the classical
Einstein equations whatsoever. Similar to the vast number of 
proposals from cosmologists, particle physicists and string theorists 
for possible 
extensions of the field content of the standard model such as inflatons,
axions, dark matter, supersymmetric extensions, dilatons, Kaluza Klein 
modes etc. 
we consider the existence of a  
fundamental dust field in addition to standard matter as an at least
theoretical possibility which of course has to be tested in experiments.
The fact that the dust only interacts gravitationally makes it 
in principle a perfect dark matter candidate\footnote{One could call it 
a NIMP
(Non Interacting Massless Particle) as compared to a WIMP (Weakly 
Interacting Massive Particle) which is considered as the most favourite
(cold) dark matter candidate. Massless here refers to the fact that the 
dust Lagrangian does not contain a usual mass (or potential) term. 
However,
the foliation defined by the dust is always spacelike and the 
corresponding foliation vector field has unit timelike norm with respect
to the physical metric (but it is not necessarily normal to the 
foliation). That is, the dust moves along unit timelike
geodesics which also follows from the fact that the dust Lagrangian
can be interpreted as the sum (or integral) of relativistic 
point particle 
Lagrangians (one for each flow line of the congruence) 
with variable mass depending on the flow line. This mass 
distribution is just the energy density of the dust. See 
\cite{brown_kuchar} for details. Hence the dust moves at non 
relativistic 
speeds as required by realistic dark matter models. Of course the dust
energy density has the wrong sign to explain the anomalous galactic
rotation curves but we should recall that the dust is anyway like a 
Higgs boson which disappears from the observable particle spectrum. 
There is no obstacle in adding one's favourite {\it observable} dark 
matter model.}. 

The topic addressed in this article is the spherically symmetric sector 
of the theory presented in \cite{ghtw1}. We will show in detail that a 
gauge invariant version of Einstein's equations specialised to the 
spherically symmetric case can be mapped to a family of Lemaitre -- 
Tolman -- Bondi (LTB) solutions. The only modification is that the 
equations contain a phantom dust energy density instead of the usual 
dust energy density and we will discuss explicitly which consequences 
this has. 

As pointed out in \cite{thiemann1}, the existence of gauge invariant 
version of Einstein's equations and 
thus a reduced phase space formulation of GR involving an algebra of 
observables and a physical Hamiltonian is of advantage when a 
quantisation of gravity is concerned because one sidesteps the 
difficulties involved in the anomaly free quantisation of constraints
and the construction of the physical Hilbert spaces.  
Recently, this strategy has been used in 
order to present a reduced phase space quantisation \cite{AQGIV} for 
Loop Quantum 
Gravity \cite{Rovelli,thiemannbook}. This framework developed for the 
full theory can of course be applied also to the spherically symmetric 
sector \cite{KG}.\\
\\
The paper is structured as follows:\\
\\
In section \ref{sec:observables} we review the results of 
\cite{ghtw1} for the full theory. 
In section \ref{sec:spherical_symmetry} we specialise 
the observables and the physical Hamiltonian to spherically symmetric 
space times and in section \ref{sec:eoms} derive the corresponding 
equations of motion. These equations are a gauge invariant version of 
Einstein's equations in the case of spherical symmetry. Section 
\ref{sec:solution} discusses the solution of these equations which 
belong to a family
  of LTB -- solutions. In section \ref{sec:semistatic} and 
\ref{sec:singularities} we discuss further properties of these solutions 
such as their semistatical behaviour as well as the occurrence of 
singularities. For the latter we analyse in particular the effect of the 
phantom dust energy momentum that occurs as a source in these equations. 
Finally in 
section \ref{sec:conclusions} we conclude, discuss the implications of 
the work done in this paper and give an outlook. Several appendices 
on the boundary conditions imposed on gauge invariant variables, various 
spherically symmetric coordinate systems employed in the literature and 
a 
comparison with the covariant derivation of the LTB solutions complete 
the paper.

\section{Brown -- Kucha\v{r} Dust Reduction of General Relativity} 
\label{sec:observables} 
\label{sec:relational}

In this section we summarise the analysis performed in \cite{ghtw1}
which builds on the seminal work \cite{brown_kuchar}. \\
\\
The Brown -- Kucha\v{r} dust Lagrangian is given by
\be
S_{\rm dust} = -\frac{1}{2}\int\limits_{M}d^4 X \sqrt{|\det(g)|}\rho 
[g^{\mu \nu} U_\mu U_\nu + 1] \quad , \label{dustaction}
\ee
which is coupled to the standard Einstein Hilbert action $S_{\rm EH}$ 
\be
S_{\rm EH}=\frac{1}{\kappa}\int\limits_{M}d^4 X \sqrt{|\det(g)|} 
^{(4)}R.
\ee
Standard matter Lagrangians can be added in the usual way, but
they do not couple to the dust.
Here $M$ is a four dimensional manifold which can topologically be 
identified 
with $\mathbb{R} \times \cal{X}$ for some three dimensional manifold 
$\cal{X}$ of arbitrary topology and $X \in M$ are local coordinates on 
$M$. Next, $g_{\mu \nu}(X)$  with $\mu, \nu = 0,1,2,3$ denotes a 
(Pseudo--) 
Riemannian metric on $M$. $U \in T^*M$ is a one form defined as the 
differential  $U = -dT + W_jdS^j$, $j =1,2,3$, for some scalar fields 
$T, W_j, S^j \in C^\infty(M)$. So finally the action (\ref{dustaction}) 
is a functional of $g_{\mu \nu}$ and eight scalar fields $\rho, T, W_i, 
S^i$. $^{(4)}R$ is the Ricci scalar corresponding to the metric $g$.

We need the Hamiltonian formulation of that system which can be derived 
by performing a usual ADM -- (3+1) -- split with respect to a foliation 
of 
$M\cong \mathbb{R}\times {\cal X}$. We 
denote the momentum 
conjugate to the ADM 3 -- metric $q_{ab}$ by $p^{ab}$. The momenta 
conjugate to the configuration variables $\rho,W_j,T,S^j$ are denoted by 
$Z,Z^j,P,P_j$ respectively. Latin indices range from 1 to 3. A detailed 
Dirac analysis of the occurring constraints shows that the coupled 
system is second class.
Hence, in order to proceed, one passes on to the corresponding Dirac 
bracket and solves the second class constraints explicitly. As a 
consequence it turns out that the momenta conjugate to $\rho$ and $W_j$ 
vanish and that these configuration variables can be expressed in terms 
of the remaining phase space variables. Explicitly, we have
\be
Z := 0, \qquad Z^j := 0, \qquad W_j := - \frac{P_j}{P}, \qquad \rho^2 := 
\frac{P^2}{\sqrt{\det{q}}}\Big[ q^{ab}U_a U_b + 1 \Big] \quad . 
\label{solve2ndclass}
\ee
In general, analysing second class constraint systems is a very hard 
task due to the complicated structure of the corresponding Dirac 
bracket. However, it turns out that for the system $S_{EH} + 
S_{dust}$, when restricting attention to the geometry variables 
$(q_{ab}, p^{ab})$ as well as the remaining dust variables $(S^j, P_j)$ 
and $(T,P)$, then the 
Dirac bracket in 
this sector reduces to the 
standard Poisson bracket again, which makes a further analysis of this 
system tractable.\\
We end up with a first class system possessing the following constraints
\be
c^{\rm tot} & = &  c + c^{\rm dust}  \label{tot_ham}\\
c^{\rm tot}_a & = &  c_a + c^{\rm dust}_a \label{tot_diffeo}\quad ,
\ee
where the geometry and dust contributions can be written as
\be
c_a & := & -\frac{2}{\kappa}q_{ac}D_{b}p^{bc} \\
c   & := & \frac{1}{\kappa}\frac{1}{\sqrt{\det{q}}} \big[q_{ac}q_{bd} - 
\frac{1}{2}q_{ab}q_{cd}\big]p^{ab}p^{cd} - \sqrt{\det(q)}R(q) \\
c^{\rm dust}  & := & P\sqrt{1 + q_{ab}U_a U_b} \\
c^{\rm dust}_a & := & PT_{,a} + P_j S^j_{,a} \quad .
\ee
Here $D_a$ is the covariant differential compatible with $q_{ab}$ and 
$R(q)$ denotes the Ricci scalar of $q_{ab}$. Here $P$ takes 
only non positive values and thus the energy density of the dust is 
negative or zero which is why we call it phantom dust in contrast to
\cite{brown_kuchar}. 
The 
reasons for that can be summarised as follows: If $P$ would take 
positive values then on the constraint surface we would have $c<0$.
As we will see, the derived physical Hamiltonian is approximated by 
$|c|=-c$ which in the limit of flat space would be the negative of the 
standard model Hamiltonian. One could cure this by letting time run 
backwards and defining the physical Hamiltonian by $-|c|=c$ but then 
energy would be unbounded from below. Furthermore,
the dynamical foliation generated by the 
dust would be past oriented if we chose the other sign, see \cite{ghtw1} 
for a detailed discussion. 

Concerning the interpretation of (\ref{dustaction}) we refer the reader 
to \cite{brown_kuchar}, where the authors describe that the dust action 
can actually be derived as a field theoretic generalisation of the 
concept of free massive relativistic particles moving on geodesics of 
the gravitational field created by the entire collection of particles. 
To get an idea why this interpretation is possible, one can check that 
the integral curves of $U^\mu = g^{\mu \nu} U_\nu$ describe geodesics of 
$g_{\mu \nu}$, $S^j$ is constant along each of these curves and $T$ 
describes proper time. So $S^j  = const.$ labels a geodesic and $T=  
const.$ is an affine parameter along the geodesic. This is exactly the 
reason why it so convenient to use the fields $T, S^j$ as a physical 
reference frame.

The canonical Hamiltonian that generates Hamiltonian equations for 
$(q_{ab},p^{ab}),(P,T)$ which are, together with the constraints in 
(\ref{tot_ham}) and (\ref{tot_diffeo}), equivalent to the 10 Einstein 
equations obtained in the Lagrangian framework is given by 
\be
{\bf H}_{can} := \int\limits_{\cal{X}}d^3x\big( n(x) c(x) + 
n^a(x)c_a(x)\big) \quad .
\ee
where $n$ is the so called lapse function and $n^a$ the so called shift 
vector which play the role of Lagrange multipliers.
\vspace{0.5cm}\\
Now we come to the crucial observation made by Brown and Kucha\v{r} in 
\cite{brown_kuchar} that the system $S_{EH} + S_D$ is indeed (partially) 
deparametrisable. To see this, note first, that the total Hamiltonian 
constraint (\ref{tot_ham}) can be written in equivalent form
\be
c^{\rm tot} & = & c + P\sqrt{1 + \frac{q^{ab}c^{\rm dust}_a c^{\rm 
dust}_b}{P^2}} \label{tot_ham2} \simeq c + P\sqrt{1 + \frac{q^{ab}c_a 
c_b}{P^2}} \quad .
\ee
On the constraint hypersurface the constraints $c^{\rm tot}$ and $c^{\rm 
tot}_a$ can be solved for the momenta $P$ and $P_j$ respectively and 
thus be written down in completely equivalent\footnote{By equivalence we 
mean that the constraint hypersurface generated by both constraints are 
the same.} form as
\be
\tilde{c}^{\rm tot} & = & P + h \qquad\quad h = \sqrt{c^2 - q^{ab}c_a 
c_b } \label{tot_ham3}\\
\tilde{c}^{\rm tot}_j & = & P_j + h_j \qquad h_j = S^a_j [c_a - hT_{,a}] 
\quad \label{tot_diffeo3} ,
\ee 
where we assumed that $S^j_{, a}$ is non degenerate and defined its 
inverse $S^a_j$. As shown in \cite{ghtw1} this condition is gauge 
invariant under the gauge transformations generated by (\ref{tot_ham3})
and (\ref{tot_diffeo3}).

At least the total Hamiltonian constraint is in deparametrised form now, 
because $h$ depends only on the gravitational variables $q_{ab}, 
p^{ab}$. Moreover, since the  
constraints are linear in the dust momenta, (\ref{tot_ham3}) and 
(\ref{tot_diffeo3}) form a strongly Abelian first class constraint 
algebra. As a consequence also $\{h(x), h(y)\} = 0$ because $h$ commutes 
with $P$. However, only the total Hamiltonian constraint $c^{\rm tot}$ 
is of deparametrised form but not the total diffeomorphism constraint, 
so Poisson brackets between either $h(x)$ and $h_j(y)$ or $h_j(x)$ and 
$h_j(y)$ will not vanish in general. So we achieved a partially 
deparametrised form for the coupled system of gravity and dust. Note 
that this can also be obtained when additional, for instance standard 
model, matter would be coupled to gravity and dust. The  only difference 
in this case will be that the constraints $c$ and $c_a$ above will then 
not only include gravitational contributions but consist of a sum of 
gravity plus the additional matter contributions. An example where 
additional to gravity and dust a K.G. -- scalar field was considered can 
be found in \cite{ghtw1,ghtw2}.\\
\\
We are now in the position to define distinguished coordinates 
on the reduced phase space defined by the first class system 
defined by (\ref{tot_ham3}) and (\ref{tot_diffeo3}). It is clear that
$T,S^j$ are pure gauge and that $P,P_j$ can be solved for the 
gravitational (and standard matter) field variables. Therefore it is 
natural to introduce a four parameter family of gauge fixing conditions 
defined by $T(x,t)=\tau,\;S^j(x,t)=\sigma^j$ where $\tau\in \mathbb{R}$
and $\sigma$ takes values in the dust space ${\cal S}=S({\cal 
X})$ which by assumption is diffeomorphic to $\cal X$. The dust space
labels the geodesics and $\tau$ is an affine parameter along the 
geodesics. 

The relational 
framework can now be applied and the results
can be described as follows (see \cite{ghtw1} for all the details
and the derivations):\\
Let $S^a_j(x)$ be the inverse of $S^j_{,a}(x)$ and $J(x):=\det(\partial 
S/\partial x)$. Consider
\be
\label{obs_diffall}
\tilde{q}_{ij}(\sigma) &= & \int\limits_{\chi}d^3x |J(x)| 
\delta(\sigma^j-S^j(x))S^a_i S^b_j q_{ab}\nn \\
\tilde{p}^{ij}(\sigma) &= &\int\limits_{\chi}d^3x |J(x)| 
\delta(\sigma^j-S^j(x))\frac{S^i_{,a} S^j_{,b} p^{ab}}{J}(x)\nn \\
\ee
It maybe checked explicitly that 
\be
\{ \tilde{p}^{ij}(\sigma), \tilde{q}_{kl}(\sigma')  \} = \kappa 
\delta_{(k}^i \delta_{l)}^j \delta(\sigma, \sigma') \quad .
\ee
The interpretation of (\ref{obs_diffall}) is obvious: These are the 
coordinate transformations of $(q_{ab},P^{ab})$ respectively into the 
dynamical coordinate system defined by $x_\sigma=S^{-1}(\sigma)$.
One can arrive at these expressions independently by symplectic 
reduction which explains why $(\tilde{q}_{ij},\tilde{p}^{ij})$ continue 
to be a conjugate pair. 

Next let
\be
\label{QaP}
Q_{jk}(\tau,\sigma)&:=&
\sum\limits_{n=0}^{\infty}\frac{1}{n!}
\{\tilde{h}_{\tau},\tilde{q}_{jk}\}_{(n)}\nn 
\\
P^{jk}(\tau,\sigma)&:=&
\sum\limits_{n=0}^{\infty}\frac{1}{n!}
\{\tilde{h}_{\tau},\tilde{p}^{jk}\}_{(n)}
\ee
where we introduced 
\be
\label{tilhtau}
\tilde{h}_{\tau}:=\int\limits_{\cal S}\, d^3\sigma \big(\tau - 
\tilde{T}(\sigma)\big)\tilde{h}(\sigma)\quad{\rm with}\quad 
\tilde{h}=h(\tilde{q}_{jk},\tilde{p}^{jk})
\ee
It maybe checked explicitly that (\ref{QaP}) has vanishing Poisson 
brackets with all constraints.

These expressions can no longer be described in a compact form, they
are hopelessly complicated to evaluate as functions of 
$\tilde{q},\tilde{p}$. However,
as stressed in the introduction, all we need is their Poisson algebra
and their time evolution. To that end, set 
$Q_{ij}(\sigma):=Q_{ij}(\tau=0,\sigma),\;\;
P^{ij}(\sigma):=P^{ij}(\tau=0,\sigma)$. Then it maybe checked explicitly
that 
\be
\{ P^{ij}(\sigma), Q_{kl}(\sigma')  \} = \kappa 
\delta_{(k}^i \delta_{l)}^j \delta(\sigma, \sigma') \quad .
\ee
still form a canonical pair, all other Poisson brackets vanishing.
Furthermore, let us define 
\be
H(\sigma) = \sqrt{C^2 - Q^{ij}C_i C_j}(\sigma) \quad,
\ee
where 
\be 
C:=\tilde{c}(\tilde{q}_{jk}=Q_{jk},\tilde{p}^{jk}=P^{jk}),\;\; 
C_i:=\tilde{c}_i(\tilde{q}_{jk}=Q_{jk},\tilde{p}^{jk}=P^{jk}),\;\; 
\ee
and $\tilde{c},\;\tilde{c}_i$ are just $c,c_a$ expressed in the dust
coordinate system.
Then it may be checked that the {\it physical Hamiltonian} generating 
time evolution for all 
observables $f(Q_{ij}, P^{ij})$ is given by
\be
\mathbf{H}_{\rm phys} := \int\limits_{\mathcal{S}} d^3 \sigma\,
H(\sigma) \quad . \label{h_phys1}
\ee

When looking at its variation 
\be
\delta \bf{H}_{\rm phys} 
& = & \int\limits_{\mathcal{S}} d^3\sigma\Big[ 
\Big(\frac{C}{H}\Big)\delta C - \Big(\frac{Q^{ij}C_{j} }{H}\Big)\delta 
C_{i} + \frac{1}{2 H}C_{i}C_j Q^{ik}Q^{jl}\delta Q_{ij}\Big]\\
& =: & \int \limits_{\mathcal{S}} d^3\sigma\Big( N\delta C + N^i\delta 
C_i + \frac{1}{2} H N^i N^j \delta Q_{ij}\Big) \quad ,
\ee
one sees that the physical equations of motion generated by 
$\mathbf{H}_{\rm phys}$ are almost equivalent to the ones generated by 
the canonical Hamiltonian $\mathbf{H}_{\rm can}$ with the identification 
$q_{ab}(x) \rightarrow Q_{ij}(\sigma), p^{ab}(x) \rightarrow 
P^{ij}(\sigma)$ modulo the following important differences: First, lapse 
$N$ and shift $N^i$ are not phase space independent functions 
as for $\mathbf{H}_{\rm can}$ where they only encode the arbitrariness 
of the foliation. Rather they are {\it observable} phase space functions 
composed out of the elementary fields $Q_{ij}, P^{ij}$ as
\be
N := \frac{C}{H},  \qquad N^i = - \frac{Q^{ij} C_j}{H} \quad , 
\label{lapseandshift}
\ee
Second, there is one 
additional contribution proportional to the Hamiltonian density 
$H(\sigma)$. But $H(\sigma)$ is a conserved quantity in the theory 
and 
can be freely chosen on the initial value hypersurface, so we may tune 
this term to alter the equations of motion as little as we like.

The physical Hamiltonian has an infinite number of conserved charges,
namely energy and momentum density $H(\sigma),\;C_j(\sigma)$. 
The latter ones generate {\it active} diffeomorphisms of the 
dust space
$\cal S$, they are to be considered as symmetries of the system rather
than gauge transformations generated by the {\it passive} 
diffeomorphisms of $\cal X$. Likewise, the 
former are 
related in an 
intricate way to time reparametrisation invariance in General 
Relativity. 

\section{Spherical symmetry} \label{sec:spherical_symmetry}
Now we want to specialise the general theory to spherically symmetric 
spacetimes. We will work directly at the gauge invariant level and 
assume that what one usually measures in physical experiments are not 
the gauge variant three metrics $q_{ab}$ and their canonically conjugate 
momenta $p^{ab}$ in some unphysical coordinate system $x^a$ but rather 
the {\it physical} metrics $Q_{ij}$ and their {\it physical} canonically 
conjugate momenta $P^{ij}$ measured with respect to the {\it physical 
reference frame} $\mathcal{S}$ given by the dust fields. There is 
nothing to debate about the fact that whenever we perform experiments we 
measure physical gauge invariant quantities and not the kinematical 
quantities $q_{ab}, p^{ab}$. Hence we will require spherical symmetry
with respect to the {\it physical} coordinate system $\sigma$.

Thus, spherically symmetric spacetimes $M=\mathbb{R}\times {\cal S}$ 
will 
be 
characterised by a triplet of Killing vector fields $\{\vec{\xi}_1, 
\vec{\xi}_2, \vec{\xi}_3  \}$ on $\cal S$ whose commutator algebra is 
isomorphic to 
the Lie algebra $so(3)$. As usual,
for such spacetimes it is always possible to find a 
coordinate chart in which the physical four metric $G_{\mu\nu}$ takes 
the special form
\be
G_{\mu \nu} & = &
\begin{pmatrix}
-N^2 + Q_{ij} N^i N^j & N^i\\
N^j & Q_{ij}
\end{pmatrix} \quad ,
\ee
where $i,j = 1,2,3$ and 
\be
Q_{ij}(\sr) & = & \diag \Big[\Lambda^2(\sr), \quad R^2(\sr), \quad 
R^2(\sr)\sin^2\st   \Big]\\
P^{ij}(\sr) & = & \sin\st \diag \Big[\frac{\PL(\sr)}{2 \Lambda(\sr)}, 
\quad \frac{\PR(\sr)}{4 R(\sr)}, \quad \frac{\PR(\sr)}{4 
R(\sr)}\sin^{-2}\st,   \Big] \quad ,
\ee
where $(\sr, \st, \sph)$ are spherical coordinates of a {\it physical} 
coordinate system on the spatial slices and $(\Lambda,P_{\Lambda})$ and 
$(R,P_R)$ respectively are conjugate pairs. Hence, the only non 
vanishing Poisson brackets are given by
\be
\{ \PL(\sr), \Lambda(\sr') \} = \frac{\kappa}{4 \pi} \delta(\sr, \sr'), 
\quad \{ \PR(\sr), R(\sr') \} = \frac{\kappa}{4 \pi} \delta(\sr, \sr') 
\quad .
\ee
In contrast to the gauge variant formalism lapse $N$ and shift $N^i$ are 
not arbitrary phase space independent functions but are given by 
(\ref{lapseandshift}). \\
Using these fields the geometry parts of the Hamiltonian and spatial 
diffeomorphism constraints reduce to:
\be
C & = & \frac{1}{\kappa} \sin\st \Lambda R^2 \Big[ \frac{\PL^2}{8 R^4} - 
\frac{\PL\PR}{4\Lambda R^3}  + \frac{2}{\Lambda^2}\Big( \frac{R'}{R}  
\Big)^2 + \frac{4}{\Lambda^2}\frac{R''}{R} - 
\frac{4R'\Lambda'}{\Lambda^3R}  - \frac{2}{R^2}   \Big]\\
C_\sr  & = & \frac{1}{\kappa} \sin\st \Big[ -\Lambda \PL' + R'\PR   
\Big]\\
C_\st  & = & C^{geo}_\sph   =  0 \quad ,
\ee
Hence, the  physical Hamiltonian ${\bf H}_{\rm phys}$, which generates 
the evolution of observables, specialised to the case of spherically 
symmetric spacetimes reads as
\be
\bf{H}_{\rm phys} & = & \int \limits_{\mathcal{S}} d^3 \sigma\, 
\sqrt{C^2 - \frac{1}{\Lambda^2}C_\sr^2} \quad . \label{Hphyssph}
\ee
\section{Equations of motion} \label{eoms} \label{sec:eoms}

Now we want to discuss the physical equations of motion for the 
spherically symmetric case: Physical time evolution is generated by the 
physical Hamiltonian (\ref{h_phys1}), so as a first step we need to 
compute its variation. This was already demonstrated for the general 
case at the end of section \ref{sec:relational} and now we want to 
specialise this result to the spherically symmetric case: For this 
purpose we must use the physical Hamiltonian ${\bf H}_{\rm phys}$ that 
has been specialised to the spherically symmetric case in 
(\ref{Hphyssph}) and compute its variation while considering $\Lambda, 
R, \PL$ and $\PR$ as the basic variables. This results in the following 
first order equations of motion:
\be
\dot{\Lambda} = \big\{{\bf H}_{\rm phys}, \Lambda \big\}   & \qquad & 
\dot{R} = \big\{{\bf H}_{\rm phys}, R \big\}  \nn \\ 
\dot{\PL} =  \big\{{\bf H}_{\rm phys}, \PL \big\}   & \qquad & \dot{\PR} 
= \big\{{\bf H}_{\rm phys}, \PR \big\} \quad .
\ee
The explicit derivation of physical equations of motion from the 
variation $\delta {\bf H}_{\rm phys}$ of the physical Hamiltonian has 
been done for full GR in \cite{ghtw1} and it was shown that suitable 
boundary conditions can be chosen so that boundary terms can be 
neglected in that calculation. This carries over to the spherically 
symmetric case. A discussion of boundary conditions in the latter case 
is given in appendix \ref{sec:boundary}.
Denoting derivatives with respect to $\tau$ and $\sr$ with dot and slash 
respectively the first order equations of motion explicitly read as
\be
\dot{\Lambda} & = & \frac{N \Lambda \PL}{4 R^2} - \frac{N \PR}{4 R} + 
(N^\sr)'\Lambda + N^\sr \Lambda' \label{Ldot} \\
\dot{R} & = & - \frac{N \PL}{4 R} + N^\sr R' \label{Rdot}\\
\dot{\PL} & = & -\frac{N C(\sr)}{4 \pi \Lambda} - \frac{N \PL \PR}{4 
\Lambda R} - \frac{4 N' R R'}{\Lambda^2} + \frac{4 N R'' R}{\Lambda^2} - 
\frac{4 N R R' \Lambda'}{\Lambda^3} \nn \\
& & + N^\sr \PL' - (N^\sr)^2 \Lambda H \label{PLdot} \\
\dot{\PR} & = & - \frac{2 N C(\sr)}{4 \pi R} + \frac{N \Lambda \PL^2}{2 
R^3} - \frac{3 N \PL \PR}{4 R^2} + \frac{4 N R^{'2}}{\Lambda R} + 
\frac{4 N R''}{\Lambda} - \frac{4 N R' \Lambda'}{\Lambda^2} \nn \\
& & - \frac{4 N \Lambda}{R} - \frac{4 N'' R}{\Lambda} - \frac{4 N' 
R'}{\Lambda} + \frac{4 N' R \Lambda'}{\Lambda^2}  + (N^\sr)' \PR + N^\sr 
\PR \label{PRdot} \quad .
\ee
We obtain the same equations when we specialise the first order 
Hamiltonian equations for full GR given in \cite{ghtw1} to spherical 
symmetry and additionally set the K.G. -- scalar field contributions to 
zero.
Formally these equations of motion coincide with the ones generated by 
${\bf H}_{\rm can}$ for the gauge variant 3--metrics and momenta up to 
the term proportional to the Hamiltonian density $H$ in (\ref{PLdot}), 
but there is one important difference:  $N$ and  $N^\sr$ are no free 
functions anymore. Rather they are phase space dependent functions given 
by (\ref{lapseandshift}) and choosing different $N, N^i$ is equivalent 
to working in a different {\it physically distinguishable} coordinate 
system. This again is due to the fact that Diff($\mathcal{S}$) is the 
group of {\it active diffeomorphisms on} $\mathcal{S}$ as opposed to the 
passive diffeomorphism on $\cal{X}$.
\\
The Hamiltonian density $H(\sigma)$ as well as  $C^{\sr}$ are constants 
of motion \cite{ghtw1}. The latter corresponds to the momentum density 
of the dust denoted by $\epsilon^{\sr}(\sigma)$, that is 
$C^{\sr}(\sigma)=-\epsilon^{\sr}(\sigma)$. In the following we will 
restrict our discussion to the case of vanishing dust momentum density. 
Then the shift vector $N^\sr$ being proportional to $C^\sr$ is 
vanishing. Furthermore the Hamiltonian density is given by $H = 
\sqrt{C^2 - Q^{ij}C_iC_j} = C$ and we automatically get unit lapse 
$N=1$. This means that equations (\ref{Ldot}) - (\ref{PRdot}) simplify 
significantly, all terms proportional to $N^\sr, N'$ and $N''$ vanish. 
Using that $H(\sigma)$ is a constant of motion, meaning that it does not 
evolve in physical time $\tau$, we can write
\be
H(\sigma) & = & \sin\st \epsilon(\sr) \qquad \mbox{for } \epsilon(\sr) > 
0 \quad .
\ee 
From $H(\sigma)=-P(\sigma)=\rho_{\rm dust}\sqrt{\det(Q)}$ we obtain
\be
\rho_{\rm dust}(\tau, \sr) & = & - \frac{\sin\st 
\epsilon(\sr)}{\sqrt{\det Q}} = - \frac{\epsilon(\sr)}{\Lambda(\tau, 
\sr) R^2(\tau, \sr)} \label{rhoD} \quad .
\ee
Hence, we can write
\be
C(\sr) := \int d\st d\sph C(\sigma) = - \int d\st d\sph \sin \st 
\Lambda(\sr)R^2(\sr) \rho_D(\tau, \sr) = -4 \pi \rho_D(\sr) 
\Lambda(\sr)R^2(\sr) \label{Crho} \quad .
\ee
For further analysis we want to transform the first order system 
(\ref{Ldot}) - (\ref{PRdot}) into a second order system. To this end one 
can solve (\ref{Ldot}) and (\ref{Rdot}) for the momenta and then use 
these expressions and their $\tau$-derivatives in (\ref{PLdot}) and 
(\ref{PRdot}). Setting $N=1$ and $N^\sr = 0$  and using (\ref{Crho}) 
this results in
\be
\frac{\ddot{\Lambda}}{\Lambda}  +  \frac{\ddot{R}}{R}  -  
\frac{\dot{\Lambda} \dot{R}}{\Lambda R}  -  \frac{\Lambda' R'}{\Lambda^3 
R}  -  \Big( \frac{\dot{R}}{R}  \Big)^2  +  \frac{1}{\Lambda^2} \Big(  
\frac{R'}{R} \Big)^2  +  \frac{2R''}{\Lambda^2 R}  -  \frac{1}{R^2} & =  
& -\frac{\rho_D}{2} \quad , \label{2ndI} \\
\frac{2 \ddot{R}}{R}  +  \frac{2 R''}{\Lambda^2 R}  -  \frac{2 
\dot{\Lambda} \dot{R}}{\Lambda R} - \frac{2 \Lambda' R'}{\Lambda^3 R} & 
= & -\frac{\rho_D}{2} \label{2ndII} \quad .
\ee
Furthermore we can use that $H(\sigma)=-\rho_{\rm dust}\sqrt{\det(Q)}$ 
yielding
\be
-\Big( \frac{\dot{R}}{R} \Big)^2 + \frac{1}{\Lambda^2} \Big( 
\frac{R'}{R}  \Big)^2  -  \frac{2 \dot{R} \dot{\Lambda}}{\Lambda R}  -  
\frac{2 R' \Lambda'}{\Lambda^3 R}  +  \frac{2 R''}{\Lambda^2 R}  -  
\frac{1}{R^2}  & = & - \frac{\rho_D}{2} \label{2ndIII} \quad ,
\ee
and the vanishing of the momentum density
\be
\dot{R}'  -  \frac{\dot{\Lambda}}{\Lambda}R' = 0 \label{2ndIV} \quad .
\ee
(\ref{2ndI}) - (\ref{2ndIII}) are actually not entirely independent of 
each other, in fact it is sufficient to keep (\ref{2ndII}) and 
(\ref{2ndIII}): One can check that solutions to these two equations also 
satisfy (\ref{2ndI}).
\section{Solving the equations of motion} \label{sec:solution}
In this section we will discuss the solution of the equations of motion. 
We will mainly follow \cite{bondi} for the derivation of this solution 
using  a slightly different notation here. 
In order to solve (\ref{2ndI}) - (\ref{2ndIV}) it is first important to 
notice that the (physical $\tau$-) time dependencies of $\Lambda(\tau, 
\sr)$ and $R(\tau, \sr)$ are not independent from each other. We can 
solve (\ref{2ndIV}) and get:
\be
\Lambda(\tau,\sr)  & = & R'(\tau, \sr)\frac{1}{\sqrt{1 + E(\sr)}} 
,\label{LambdafromR}
\ee
where $E(\sr)$ is a so far arbitrary, time--independent 
function\footnote{It will become clear later on why we have chosen this 
rather complicated form for the integration constant.} of $\sr$. 
Exploiting this, we can rewrite (\ref{2ndII}) and (\ref{2ndIII}) as 
\be
\frac{2 \ddot{R}}{R}  +  \Big(  \frac{\dot{R}}{R}  \Big)^2  -  
\frac{E}{R^2} & = & 0 \label{onlyR1}\\
\frac{\ddot{R}}{R}  -  \frac{\dot{R}' \dot{R}  }{R' R}  +  \frac{E'}{2 
R' R}  & = & -\frac{\rho_D}{4} \quad . \label{onlyR2}
\ee
First, by multiplying (\ref{onlyR1}) by $R^2 \dot{R}$ we get
\be
2 R \dot{R} \ddot{R} + \dot{R}^3 -E\dot{R} = R(2 \dot{R} \ddot{R} ) + 
\dot{R} (\dot{R}^2 - E  ) = \frac{d}{d \tau} \Big[ R(\dot{R}^2 - E)  
\Big] & = & 0 \nn \\
\Rightarrow \dot{U} & = & 0 \quad ,
\ee
where we defined $U = R (\dot{R}^2 - E)$. $U$ does not depend on $\tau$ 
so we can write $U(\tau, \sr) = F(\sr)$ for an so far arbitrary 
$\sr$-dependent function $F(\sr)$. By differentiating $U$ with respect 
to $\sr$ we get
\be
U'= F' = R' (\dot{R}^2 - E) + 2 R \dot{R} \dot{R}' - E' R  \quad .
\ee
Dividing this by $R^2 R'$ we see that $F'$  is equal to the lefthand 
side of (\ref{onlyR1}) minus 2 times the lefthand side of  
(\ref{onlyR2}). Thus, we can rewrite $F'$ as
\be
F' & = & \frac{1}{2}R' R^2 \rho_D = \frac{1}{2}\sqrt{1 + E}\,\Lambda R^2 
\rho_D = - \frac{1}{2}\sqrt{1+E}\, \epsilon(\sr) \quad ,
\ee
where we used (\ref{LambdafromR}) in the first equality and (\ref{rhoD}) 
in the second one. 
Integrating the equation above once, we end up with
\be
\label{DefF}
F(\sr) & = &-\frac{1}{2} \int\limits_{0}^{\sr} d\lambda\, 
\epsilon(\lambda)\sqrt{1 + E(\lambda)} + \alpha =: - \frac{1}{2} M(\sr) 
+ \frac{1}{2} M(0) \quad ,
\ee
where $M(0)$ is an so far arbitrary (constant) mass. $M(\sr)$ can be 
interpreted as the effective gravitating mass (multiplied by 1/4 since 
the relation between the Schwarzschild radius and the central mass 
involves a factor of 2) of the dust inside a sphere with radial label 
$\sr$. So finally the equation $U = F$ yields
\be
\dot{R}=\pm\sqrt{\frac{F}{R}+E}\quad . \label{Newton1}
\ee
We arrive at the following general form of the metric
\be 
\label{solEOMgen}
ds^2 & = & -d\tau^2 + \frac{R^{'2}(\tau, \sr)}{1 + E(\sr)}d\sr^2 + 
R^2(\tau, \sr)d\Omega^2 \nn \\
\dot{R}(\tau, \sr) & = & \pm \sqrt{E(\sr) + \frac{F(\sr)}{R(\tau, \sr)}  
} \nn \\
\rho_D(\tau, \sr) & = &  \frac{2 F'(\sr)}{R'(\tau, \sr) R^2(\tau, \sr)} 
\quad ,
\ee
with $E(\sr)> -1$ and $F(\sr)$ must be chosen such that the expression 
under the square root in the second line is non--negative for all 
$R(\tau, \sr)$, see section \ref{sec:singularities} for further 
elaborations on this topic.\\
The solution for $R(\tau,\sr)$ for general $E(\sr)$ can only be given in 
parametric form, distinguishing the cases $E>0, E=0, E<0$ and using the 
conventions of \cite{landau_lifshitz} one finds:
\begin{itemize}
\item{$E(\sr)>0:$}
\\
\be
\label{EgNull}
R(\tau,\sr)& = &\frac{F(\sr)}{2E(\sr)}\Big({\rm cosh}(\eta) - 1\Big)\nn 
\\
\Big({\rm sinh}(\eta) - \eta\Big) &=& 
\frac{2\big[E(\sr)\big]^{\frac{3}{2}}(\beta(\sr) - \tau)}{F(\sr)}
\ee
\item{$E(\sr)<0:$}
\\
\be
\label{EkNull}
R(\tau,\sr)& = &\frac{F(\sr)}{2(-E(\sr))}\Big(1 - {\rm cos}(\eta) 
\Big)\nn \\
\Big(\eta - {\rm sin}(\eta) \Big) &=& 
\frac{2\big[-E(\sr)\big]^{\frac{3}{2}}(\beta(\sr) - \tau)}{F(\sr)}
\ee
\item{$E(\sr)=0:$}
\\
\be
\label{lem+1}
R(\tau,\sr)& = & \Big[\frac{3}{2}\sqrt{F}\big(\beta(\sr) - \tau 
\big)\Big]^{2/3}
\ee
\end{itemize}
Here $\beta(\sr),E(\sr)$ and $F(\sr)$ are so far arbitrary functions of 
$\sr$ which allow a coordinate choice by a particular form of 
$\beta(\sr)$ and two further physical quantities. As mentioned above $F$ 
in (\ref{DefF}) can be understood as the effective gravitating mass 
within radius $\sr$  and $E(\sr)$ determines the time evolution of $R$ 
as well as the local geometry. For a more detailed discussion see for 
instance \cite{szekeres_lun}. The time $\tau$ at which $R(\tau,\sr)$ is 
equal to zero is $\beta(\sr)$ and one calls $\tau\ge\beta(\sr)$ the big 
bang time whereas $\tau\le\beta(\sr)$ is referred to as the recollapse 
time.
Often one refers to these three cases as elliptic, parabolic and 
hyperbolic for $E(\sr)<0, E(\sr)=0$ and $E(\sr)>0$ respectively.
The solution in (\ref{solEOMgen}) can easily be identified as the whole 
class of Lema\^{i}tre--Tolman--Bondi (LTB) metrics (see appendix 
\ref{LTB}), so we managed to map the problem of solving the physical 
$\tau$-evolution of gauge--invariant three--metrics $Q_{ij}$ to the 
problem of analysing the LTB class of solutions in standard GR. However, 
it is important to note that opposed to the standard LTB models, the 
mass term $M(\sr)$ enters with the opposite sign in $F(\sr) = 
\frac{1}{2}(M(0) - M(\sr))$. That means the effective gravitating mass 
will decrease when going further away from the central mass. This 
happens because the dust clocks, which have been chosen as a physical 
reference system, have negative energy.\\
\\
We can fix the coordinate choice included in $\beta(\sr)$ and the mass 
$M(0)$ in (\ref{DefF}) by requiring that in the vacuum case the general 
solution above should reduce to the Lema\^{i}tre solution 
\cite{lemaitre} given by
\be
ds^2&=&-d\tau^2+\frac{R_s}{R_{\rm 
Lem}(\tau,\sr))}d\sr^2+R_{{\rm Lem}}^2(\tau,\sr)d\Omega^2\quad{\rm 
with}\quad 
R_{\rm Lem}(\tau,\sr)=\Big[\frac{3}{2}\sqrt{R_s}(\sr - \tau)\Big]^{2/3}
\ee
where $R_s=2MG/c^2$ denotes the Schwarzschild radius which simplifies in 
units $h=c=G=1$ to $R_s=2M$.
The Lema\^{i}tre solution is diffeomorphic to the Schwarzschild solution 
and can be obtained from the latter through a coordinate transformation 
into a comoving coordinate system. So this metric describes the local 
coordinate system of a free falling observer in a spherically symmetric 
gravitational field originating from a central mass M, more details can 
be found in appendix \ref{lemaitre}. 
Matching the Lemaitre with the LTB solution we obviously need to 
consider the 
case $E(\sr)=0$. 
This is 
often referred to as the marginally bound case. Furthermore, in the  
vacuum case,
$\epsilon(\sr)$ is zero and as a consequence $F^{\prime}$ vanishes. 
Hence, in this case $F$ is just equal to the constant $\frac{1}{2}M(0)$ 
introduced in (\ref{DefF}). Using this we obtain 
$R(\sr,0)=\big[3/2\sqrt{\frac{1}{2}M(0)}(\beta(\sr))\big]^{2/3}$ whereas 
$R_{\rm Lem}(\sr,0)=\big[3/2\sqrt{R_s}(\sr)\big]^{2/3}$. Thus, we can 
match these two radii by simply choosing $\beta(\sr)=\sr$ and 
identifying $M(0) = 4M$ because $R_s=2M=1/2 M(0)$.
\subsection{Newtonian limit}
Let us have a look at the Newtonian limit of this spacetime: In 
\cite{bondi} it was argued that $R(\tau, \sr)$ is the quantity that 
should actually be identified with ``Newtonian distance'', for instance 
in the context of luminosity distance. Having this in mind let us look 
at equation (\ref{Newton1}) again. By differentiating this one with 
respect to $\tau$ we obtain
\be
\ddot{R} = - \frac{F}{2 R^2} \quad ,
\ee
which coincides formally with Newton's equation of motion for a point 
particle under the influence of a central mass. The crucial difference 
is that the effective gravitating mass $F/2$ is less than the central 
mass $M=R_s/2$. So the gravitational field is weaker than one would 
expect without having the dust around.\\
There is one remark necessary concerning the falloff behaviour of 
$R(\tau, \sr)$: One might wonder whether the boundary conditions 
discussed in appendix \ref{sec:boundary} are not violated by the 
explicit form of $R$ as in (\ref{lem+1}). The solution to this puzzle 
lies in the fact that in the derivation of the falloff conditions 
(\ref{asymptotics}) we assumed a coordinate system which approaches a 
flat spherical Minkowskian one in the asymptotic limit. In contrast, the 
coordinate system we are using here is the coordinate system of a freely 
falling observer so one cannot expect (\ref{lem+1}) to hold 
automatically.{
 Nevertheless, we assumed that $\epsilon(\sr) \rightarrow 0$ for $\sr 
\rightarrow \infty$, so in the limit of vanishing dust density and in 
the marginally bound case (that is E=0) the LTB coordinates merge into 
Lemaitre coordinates which themselves can be transformed to 
Schwarzschild coordinates having the correct Newtonian limit (see 
appendix \ref{lemaitre}) .
\\
\\
We have seen above that the exact Schwarzschild solution is obtained 
for $F=$const. and $E=0$. However, it is possible to obtain the 
Schwarzschild solution also for certain $E\not=0$ and $F=$const. The 
idea 
is to transform the LTB coordinates first into 
generalised Painleve -- Gullstrand coordinates (GPG), 
introduced in \cite{LaskyLunBurston} and then to Schwarzschild 
coordinates. For later comparison, let us first 
transform the Lemaitre solution from Lemaitre
coordinates $(\tau,\sigma_r)$ to the ordinary 
Painleve -- 
Gullstrand (PG) coordinates $(\tau,R)$ corresponding to 
observers moving on radial timelike geodesics labelled by Schwarzschild 
radial coordinate $R$. This yields a non 
-- diagonal metric whose line element is of the form 
\cite{Painleve,Gullstrand}
\be
\label{PG}
ds^2 = -d\tau^2 + \left(dR + \sqrt{\frac{R_s}{R}}d\tau\right)^2 + 
R^2d\Omega^2
\ee
where we used $\tau$ for the Lemaitre and PG time coordinate. We present 
in  appendix \ref{GPsol} how to transform from Schwarzschild to PG 
coordinates. For a pedagogical introduction to PG coordinates and more 
details see for instance \cite{MartelPoisson}. Often one uses Kruskal 
coordinates to obtain a maximally extended Schwarzschild spacetime. The 
PG coordinates do not cover the total Kruskal manifold. In the form 
given in equation (\ref{PG}) they cover the black hole horizon and the 
black hole singularity at $R=0$ hence the physically interesting part of 
the manifold for our discussion. If we choose the sign of 
$\dot{R}=\pm\sqrt{E+F/R}$ to be negative then we obtain another set of 
coordinates with a line element similar to that in (\ref{PG}) but with a 
minus sign in front of the off diagonal term. These PG -- coordinates 
will then describe the so called "white hole region" of the Kruskal 
manifold. One of the properties of Kruskal coordinates is that $R$ is 
given only implicitly therefore for our purpose of discussing the 
Newtonian limit GPG coordinates are more appropriate. One of the 
interesting properties of the ordinary PG metric shown in (\ref{PG}) is 
that the spatial hypersurfaces (that is $d\tau=0$) are flat since then 
$ds^2=dR^2+R^2d\Omega^2$ and all  information about the curvature is 
encoded in the shift vector given by $\vec{N}=(N^R:=\sqrt{R_s/R},0,0)$. 
Furthermore, asymptotically ($R\to\infty$) the PG metric coincides with 
the Minkowski metric.
\\
\\
Now we transform the LTB coordinates 
$(\tau,\sigma_r,\theta,\phi)$ into GPG coordinates 
$(\tau,R,\theta,\phi)$. Following \cite{LaskyLunBurston} this yields  
the metric:
\be
\label{GPG}
ds^2=-d\tau^2 +\frac{ \left(dR 
+\sqrt{E(\tau,R)+\frac{F(\tau,R)}{R}}d\tau^2\right)}{1+E(\tau,R)} + 
R^2d\Omega^2
\ee
subject to $E(\tau,R)>-1$. 
The functions $E$ and $F$ that are functions of $\sigma_r$ only in LTB 
coordinates now become functions depending on $\tau$ and $R$ in GPG 
coordinates. Therefore the time dependence of $E$ and $F$ in these 
coordinates is restricted by the partial differential equations
\begin{eqnarray}
\label{CondEF}
 \frac{\partial E(\tau,R)}{\partial \tau} -\sqrt{E(\tau,R) + 
\frac{F(\tau,R)}{R}}\frac{\partial E(\tau,R)}{\partial R}& = &0  
\nonumber \\
 \frac{\partial F(\tau,R)}{\partial \tau} -\sqrt{E(\tau,R) + 
\frac{F(\tau,R)}{R}}\frac{\partial F(\tau,R)}{\partial R} &= &0
 \end{eqnarray}
More details about the transformation from LTB to GPG coordinates can 
be found in appendix \ref{GPGSol}. We will not comment on the 
full class of solutions of this system of PDE's but just remark 
that e.g. (local) analytic solutions can 
be found by providing analytic initial data $F(0,R),\;E(0,R)$ 
and determining the coefficients of the Taylor expansion (in terms of 
$\tau$) of  
$E(\tau,R),\;F(\tau,R)$ by taking higher derivatives of (\ref{CondEF}) 
at $\tau=0$ (Kovalevskaja method). For instance, a trivial solution is 
given by taking $E=$const. and $F=$const. In the special case 
$E(\tau,R)=0$ and $F(\tau,R)=R_s=\rm{const}$ the GPG 
metric coincides with the PG metric justifying the name generalised 
Painleve -- Gullstrand metric. In contrast to the PG solution 
in general the spatial hypersurfaces of the GPG metric are no longer 
flat due to the $1/(1+E)$ factor in front of $dR^2$.
So one could think that in the limit of vanishing dust density 
$\epsilon(t,R)\to 0$ a transformation to Schwarzschild coordinates is 
only possible in the marginally bound case $E=0$. As discussed in 
\cite{LaskyLunBurston} this is not the case and we summarise their 
discussion in the following: 
\\
Let us consider the case of vanishing dust density $\epsilon(t,R)=0$.
Since $\epsilon(\tau,R)\sim F^{\prime}$ it follows   
$F=R_s$. The equation for $F$ in (\ref{CondEF}) is thus 
trivially satisfied.  For each function $E$ satisfying the equation in 
(\ref{CondEF}) there exists a transformation from GPG coordinates 
$(\tau,R,\theta,\phi)$ to Schwarzschild coordinates $(T,R,\theta,\phi)$ 
given by
\begin{eqnarray} \label{intcond}
\left(\frac{\partial\tau}{\partial T}\right)^2&=&1+E \nonumber\\
\frac{\partial\tau}{\partial 
R}&=&\frac{\sqrt{E+\frac{R_s}{R}}}{1-\frac{R_s}{R}}
 \end{eqnarray}
 Thus one obtains a family of coordinate transformations parametrised by 
the functions $E$. Indeed, the integrability condition for the system
(\ref{intcond}) leads back to the first condition in (\ref{CondEF})
(notice that one has to write $E(T,R)\equiv E(\tau(T,R),R)$ in order to 
derive it).
 \\
 Therefore by first transforming the LTB to GPG coordinates it is 
possible to transform to Schwarzschild coordinates in the limit of 
vanishing dust density for all (allowed) functions $E$. Thus we can 
transform to Schwarschild coordinates not only in the marginally bound 
case $E=0$ but also in the elliptic $(E<0)$ and hyperbolic $(E>0)$ 
cases provided $F=$const. 
Consequently, the Newtonian limit is correctly implemented for all 
values of $E$ and $F=$const.
}

\section{Semistatic properties} \label{sec:semistatic}
Now we want to analyse whether the physical spacetime $(M,G)$ under 
consideration is semistatic in a certain sense or not. We already know 
that the physical metric $G_{\mu \nu}$ takes the form of a 
Lema\^{i}tre metric in the limit $\epsilon(\sr) \rightarrow 0$ which in 
turn is just a Schwarzschild spacetime written in comoving coordinates 
(see appendix \ref{lemaitre}). So one would expect to recover its static 
properties in some limit.\\
Let us have a look at the Killing equation
\be
(\mathcal{L}_{\vec{\xi}} G_{\mu \nu}) \equ 0 \label{killing}
\ee
for the observable metric $G_{\mu \nu}$ and examine whether there 
exists a timelike Killing vector field $\vec{\xi}$. For the Lema\^{i}tre 
metric such a Killing vector field is given by $\vec{\xi} = 
[1,1,0,0]^T$, at least outside the event horizon, so we would expect to 
find something similar here. We will start with the ansatz
\be
\vec{\xi} & = & \Big[ \xi^\tau(\tau, \sr), \xi^\sr(\tau, \sr), 0, 0   
\Big]^T \quad .
\ee
For the physical metric described above (\ref{killing}) reduces to
\be
\dot{\xi}^\tau & = & 0 \label{kill1} \\
\xi^{\tau}\dot{R'} + \xi^\sr R'' + R' ({\xi^\sr})^{\prime} & = & 0 
\label{kill2}\\
\xi^\tau \dot{R} + \xi^\sr R' & = & 0 \label{kill3}\\
-(\xi^\tau)^{\prime} + {R'}^2\dot{\xi}^\sr & = & 0 \quad . \label{kill4}
\ee
From (\ref{kill3}) we get
\be
\xi^\sr & = & -\frac{\dot{R}}{R'}\xi^\tau \quad , \label{kill5}
\ee
and using (\ref{kill1}) we obtain
\be
\dot{\xi}^\sr & = & \Big[ -\frac{\ddot{R}}{R'} + 
\frac{\dot{R}\dot{R}'}{{R'}^2}  \Big]\xi^\tau \quad .
\ee
(\ref{kill4}) can be solved for
\be
 (\xi^\tau)^{\prime} & = & \Big[ -R' \ddot{R} + \dot{R} \dot{R}' 
\Big]\xi^\tau \quad,
\ee 
and using this in (\ref{kill5}) we see that
\be
(\xi^\sr)^{\prime} & = & \Big[ -\frac{\dot{R}'}{R'}  +  
\frac{\dot{R}R''}{{R'}^2} + \dot{R}\ddot{R} - 
\frac{{\dot{R}}^2\dot{R}'}{R'}  \Big]\xi^\tau \quad .
\ee
Putting all this together we obtain a condition on the solution
\be
R' \ddot{R} - \dot{R}\dot{R}' & \equ & 0 \quad ,
\ee
and using the explicit form of the solution (\ref{lem+1}) we see that 
this holds only if
\be
F'(\sr) & \equ & 0 \quad .
\ee
This means that there exists a timelike Killing vector field $\vec{\xi}$ 
only if $\epsilon(\sr) = 0$, that is if the dust matter fields are 
vanishing.\\
Nevertheless, as one can easily check, the metric is invariant under the 
action of the vector field
\be
\vec{\xi}_0 & := & \Big[1, -\frac{\dot{R}}{R'}, 0, 0   \Big]^T
\ee
up to terms of at least $\mathcal{O}(\epsilon(\sr))$. So with
\be
(\mathcal{L}_{\vec{\xi}_0} G)_{\mu \nu} & = & 
\mathcal{O}(\epsilon(\sr))
\ee
there is a precise sense in which the physical metric $G_{\mu \nu}$ 
can be called ``semistatic''. For vanishing $\epsilon(\sr)$ this would 
be an exact symmetry and the metric would be isomorphic to the standard 
Schwarzschild metric. $\epsilon(\sr) = 0$ would be inconsistent  because 
we are real dust fields as dynamical clocks rather than ideal ones and 
vanishing 
$\epsilon(\sr)$ 
would mean that the clocks do not carry any energy at all\footnote{The 
dust part of the Hamiltonian constraint would vanish.}. Nevertheless, 
$\epsilon(\sr)$ is a free function in the theory and we can choose it 
arbitrarily small\footnote{There are certain restrictions on 
$\epsilon(\sr)$ which have to be fulfilled in order to avoid 
singularities, see section \ref{sec:singularities} for further comments 
on this issue.}. So by tuning $\epsilon(\sr)$ in the right way we can 
get as close to a static spacetime as we wish.

\section{Discussion of Singularities} \label{sec:singularities}
In this section we will discuss the possible singularities which might 
appear for the metric given in (\ref{solEOMgen}) and displayed again 
below: 
\be
ds^2 &  = & -d\tau^2 + \frac{{R'}^2}{1 + E(\sr)}d\sr^2 + R^2d\Omega^2 
\quad .
\ee
As is well known from earlier studies concerning LTB spacetimes (see for 
example \cite{hellaby_lake, szekeres_lun}) there appear two different 
kinds of singularities: Recalling that the dust density was given by 
$\rho_{\rm dust} = \frac{F'}{R' R^2}$ there are potentially two cases 
when it can diverge.
The first one, if $R(\tau, \sr) = 0$, is called a {\it collapse 
singularity} in the literature, and the second, if $R'(\tau, \sr) = 0$, 
is known as a {\it shell crossing singularity}. It was pointed out by 
\cite{zeldovich_grishchuk} for the first time that in the latter case an 
appropriate decay behaviour for $F'$ when $\sr\to\infty$ can avoid the 
divergence of $\rho_{\rm dust}$ in the shell crossing case. From the 
line element above we can read off that in the case of a collapse 
singularity the metric components $Q_{\theta\theta}$ and $Q_{\phi\phi}$ 
vanish while for a shell crossing singularity only $Q_{\sr\sr}$ is zero.
\subsection{Shell crossing singularities}
For standard LTB spacetimes with positive dust densities one can always 
choose the arbitrary functions $E(\sr), F(\sr)$ in such a way that shell 
crossing singularities will not appear \cite{hellaby_lake}. This holds 
not only for the marginally bound case ($E(\sr) = 0$) but also for the 
hyperbolic ($E(\sr)> 0$) and elliptic ($-1 < E(\sr) < 0 $) case.\\
In our framework the dust energy density enters with negative sign.  
Therefore $F'(\sigma)$ has negative sign. Furthermore we have the 
requirement  $F/R + E > 0$. Consequently we arrive at slightly different 
conditions on $E(\sr), F(\sr)$ in order to avoid shell crossing 
singularities.\\
The analysis for $E(\sr)\not=0$ involves the parametric solutions for 
$R(\tau,\sr)$ shown in (\ref{EgNull}) and (\ref{EkNull}). A singularity 
discussion becomes more involved in these two cases and we will restrict 
ourselves to the marginally bound case ($E(\sr)=0)$ here, similar 
results can also be obtained for the hyperbolic and elliptic case.\\
If $E=0$ then $Q_{\sr \sr}$ is given by
\be
{R'}^2(\tau,\sr) & = & \frac{1}{4 R(\tau,\sr) F(\sr)} \Big[ F'(\sr) (\sr 
- \tau) + 2 F(\sr) \Big]^2 \quad.
\ee 
Considering the constraints $F>0,\;F'<0,\;\sigma_r-\tau\ge 0$, in order
to avoid shell crossing, i.e. $R'\not=0$ for all $\tau,\sigma_r$ we 
obviously must have 
\be 
0<-F' < \frac{2F}{\sr-\tau} \;\;{\rm or}\;\; -F' > \frac{2F}{\sr-\tau}>0 
\ee
for all $\sr,\tau$.
Since $F$ only depends on $\sr$, for any given $\sr$ we can choose 
$\sr-\tau$ arbitrarily small so that for any choice of $F$ the second 
possibility can be violated. Hence the only dynamically stable condition 
is 
\be 
0<-F' < \frac{2F}{\sr-\tau} 
\ee
Since $\sr\ge \sr-\tau\ge 0$, this condition is certainly implied by
\be 
0<-F' < \frac{2F}{\sr} \;\; \Leftrightarrow \;\; 0<[\ln(F\sr^2)]' 
\ee
Since the logarithm is an increasing function, we obtain that $F\sr^2$
should be an increasing function while $F$ should be a decreasing 
function. This leaves us with a large class of possible $F$, for 
instance $F(\sr)=\frac{R_s}{1+\sr}$ which also fulfills $F\le R_s$.
Inserting this form of $F$ into $R$ 
and $R^{\prime}$ yields the following density for the dust
\be
\rho_{\rm dust} = -\frac{2}{3}\frac{1}{\left(1+\frac{1}{2}\left(\sr + 
\tau\right)\right)\left(\sr - \tau\right) }
\ee
The dust density diverges at $\tau=\sr$ as expected because this is 
exactly the singularity at $R=0$ that can not be avoided, see also 
discussion in the next section. If we take the limit $\sr\to\infty$ for 
fixed values of $\tau$ then $\rho_{\rm dust}\to 0$. Hence, $\rho_{{\rm 
dust}}$ shows a physically reasonable behaviour.

\subsection{Collapse singularities}
The second kind of singularities are those where $R=0$ and therefore 
besides the dust density $\rho_{\rm dust}$ also the expression for 
$\dot{R}^2$ in (\ref{solEOMgen}) becomes singular. These singularities 
occur for all three cases --  marginally bound, hyperbolic and elliptic 
-- when $\tau \rightarrow \sr$. Having collapsing dust shells in mind 
this is exactly the moment when the dust shell labelled by $\sr$ reaches 
the singularity. Note that close to $R=0$ the expression for 
$\dot{R}^2$ is dominated by the $E=0$ regime and thus we will again 
restrict our discussion to the marginally bound case. Looking  at the 
explicit form of $R(\tau, \sr)$ written down in (\ref{lem+1}) we can 
analyse how the different sign of the dust alters the behaviour during 
the collapse. In contrast to the shell crossing singularity there is no 
way to get around this collapse singularity for the system gravity plus 
pressureless dust.
As long as we do not introduce additional matter fields and take their 
non--gravitational interactions into account\footnote{In 
phenomenological matter models this is usually done by introducing 
matter with non vanishing pressure. This pressure is mandatory in order 
to describe stable configurations such as stars. In the gauge invariant 
formalism discussed in this article one would essentially follow the 
same steps, but one has to take into account the additional component 
$\rho_{\rm dust}$ when setting up the equation of state. For 
non--collapsing configurations $\rho_{\rm dust}$ could be tuned 
arbitrarily small.} we end up with a singularity after $\tau = \sr$, 
just as in the gauge variant framework using standard LTB -- models.\\
\vspace{0.5cm}\\
Finally we want to summarise all the physical selection criteria for 
$\epsilon(\sr)$ in the marginally bound case we have encountered so far: 
First, due to the special role of the dust fields in our formalism, we 
demand $F' <0$ or equivalently $\epsilon > 0$. Second we need to ensure 
that $0 < F(\sr) \leq R_s$, otherwise (\ref{lem+1}) is not a solution 
anymore; in terms of $\epsilon$ this means that 
$\int\limits_{0}^{\infty} d \sr \epsilon(\sr)< 2R_s$. Third, we need to 
choose $\epsilon(\sr)$ such that $|F'(\sr)|=-F'(\sr)  <  \frac{2 
F(\sr)}{\sr}$ is 
fulfilled in order to not run into any shell crossing singularities. 
Given these conditions the physical metric $G_{\mu \nu}$ is well 
defined until $\tau \rightarrow \sr$ when a free falling observer 
approaches the central singularity.
\section{Conclusions} \label{sec:conclusions}{
The task  of finding Dirac observables for General Relativity involves 
the construction of quantities that commute with all constraints of 
General Relativity. Dirac observables are thus the associated to gauge 
invariant objects for General Relativity where the gauge group is 
closely related to ${\rm Diff}({\cal M})$, the group of diffeomorphism 
of 
the underlying manifold ${\cal M}$. The reason why it is more 
complicated to construct gauge invariant quantities in the framework of 
General Relativity is that the mathematical structure  of ${\rm 
Diff}({\cal M})$ is richer than the structures of the gauge groups used 
for instance in the standard model of particle physics. 
In the context of the Relational formalism one can at least formally 
construct Dirac observables , with respect to chosen clocks, one for 
each occurring constraint. Choosing these clocks means choosing an 
observer that is dynamically coupled to the system. Considering the case 
of vacuum gravity these clocks must necessarily be four components of 
the four metric $g_{\mu\nu}$. If one applies the techniques developed in 
the relational framework and tries to compute (Dirac) observables for 
pure GR then one realises that the physical time evolution of those 
observables does not resemble the Einstein equations. This is because  
their 
dynamics is described by an observer who sits in a laboratory whose 
motion through spacetime 
is {defined through} the dynamics of these four metric components. 
Of course, theoretically there always exist  a coordinate transformation 
from this observer to for instance a free falling observer, however, 
practically such a transformation will be hard to find.
Fortunately, 
for certain types of matter coupled to gravity, the constraints, 
consisting of the gravitational and the matter contribution, can be 
rewritten in (partially) deparametrised form. Hence, these enlarged 
systems fall into the class of {\it deparametrisable theories} for which 
the question of observables can be addressed technically easier then in 
the general case. 
This idea was used in \cite{ghtw1, ghtw2} where instead of considering 
pure gravity (i.e. the Einstein--Hilbert action) one considers gravity 
plus pressureless dust--matter fields (i.e. Einstein--Hilbert plus dust 
action). The dust fields become the clocks of the system and correspond 
to a dynamically coupled free falling observer.
One obtains {\it physical equations of motion} for the {\it observable 
3--metric $Q_{jk}$} and {\it observable momenta $P^{jk}$} in the 
dynamical reference frame defined through the dust fields. 
Of course, when introducing a dynamically coupled observer, we have to 
ensure 
that the occurring fingerprints of the observer are still in agreement 
with experimental data.

That this is the case for the cosmological sector was already shown in 
\cite{ghtw1, ghtw2} and in this article we demonstrated that this 
framework also describes gravitational physics in the spherically 
symmetric 
sector to arbitrary precision. We showed that within this sector the 
dynamical evolution of $Q_{ij}, P^{ij}$, generated by a physical 
Hamiltonian $\bf{H}_{\rm phys}$, is in one---to--one correspondence with 
the class of Lemaitre--Tolman--Bondi (LTB) solutions for standard GR.
Interpreted in our language, the choice of lapse and shift considered 
for the LTB solutions correponds to a gauge fixing of the (spherically 
symmetric) spacetime diffeomorphism invariance whose gauge invariant 
extension is precisely induced by our choice of clocks.
In addition, while 
in the usual LTB framework one starts with a spherically symmetric, 
pressure free, perfect fluid Ansatz for the energy momentum tensor whose 
dynamics is then derived from the Bianchi identity, here we 
start with 
a fully covariant matter Lagrangian whose equations of motion then 
reduce to the usual ones. In other words, our framework provides a 
Lagrangian underpinning of the usual LTB framework. That this works 
so well is non trivial as we discuss in appendix \ref{covariant}.
 
The LTB class of solutions has been carefully investigated in the 
literature and its dynamics {\it in the comoving frame} is well 
understood. One might object that that the LTB class does not contain 
{\it static} solutions, at least for non--vanishing dust energy which is 
mandatory if one wants to use the dust fields as clocks. But this had to 
be expected when taking into account the influence of a realistic 
observer (i.e. an observer which is {\it dynamically} coupled to the 
system as opposed to a mere test observer). In a sense one has to 
abandon the idealisation of a static (vacuum\footnote{These 
considerations do not necessarily hold for more realistic models with 
additional matter fields, i.e. equilibrium states for complex systems 
such as stars still exist, just the point of equilibrium will be 
slightly shifted due to the influence of the dust.}) spacetime when 
describing physics in terms of 
observable quantities. 

Nevertheless, we showed that there exists a well defined notion of a 
{\it semistatic spacetime} and one can get as close to the standard 
Schwarzschild solution as desired by appropriately choosing certain 
constants of motion (which are related to the dust energy density). 
Additionally, we discussed that by transforming the LTB system to 
generalised Painleve -- Gullstrand coordinates and considering the limit 
of vanishing dust density there exists a well defined coordinate 
transformation to Schwarzschild coordinates for the elliptic, hyperbolic 
and marginally bound case. Consequently, in the limit where the dust 
density can be neglected the Newtonian limit is correctly implemented.
Finally, although we consider phantom dust rather 
than usual dust, there exists a range of solutions for which 
shell crossing singularities are avoided while the collapse 
singularity is unavoidable. 

To conclude, the framework presented in \cite{ghtw1, ghtw2} seems to be 
compatible with observations, at least in the cosmological and 
spherically symmetric sectors which are the most relevant analytically 
solvable ones when it comes to phenomenological applications. 
This framework might also be useful when it comes to quantising General 
Relativity. The fact that the constraints have already been 
solved at the classical level opens the door for a reduced phase space 
quantisation as opposed to the Dirac programme which is usually employed 
in Loop Quantum Gravity \cite{Rovelli,thiemannbook}. First steps into 
this direction have already 
been performed in \cite{AQGIV} and a detailed analysis of this framework 
in the context of spherical symmetry is the subject of future work 
\cite{KG}.

$\;$\\
\vspace{1cm}
\\
{\large\bf Acknowledgements}\\
\\
{
K.G. wants to thank the Perimeter Institute for Theoretical Physics 
where part of this work was completed.
J.T. wants to thank Bianca Dittrich for discussions concerning 
different 
choices of clock variables in the framework of complete observables.
Research performed at Perimeter
Institute for Theoretical Physics is supported in part by the
Government of Canada through NSERC and by the Province of Ontario
through MRI. 
}

\appendix

\section{Boundary conditions} \label{sec:boundary}

We are dealing with asymptotically flat spacetimes and hence must 
impose suitable boudary conditions. For the full theory,
these boundary conditions for the gauge invariant observables as 
obtained 
via Brown -- Kucha\v{r} dust reduction were discussed extensively 
in \cite{ghtw1}. Here we need their reduction to spherical symmetry.
Consider, as in the main text, a spherically symmetric
coordinate system as $(\sr, \st, \sph)$. Tensor indices on the dust 
manifold $\cal S$  are denoted by 
$i, j, k,\dots = 1,2,3$. The reduced observables are 
$Q_{ij}(\sigma)$ and their canonically conjugate momenta 
$P^{ij}(\sigma)$, both of which are {\it observables} in the sense 
described above. As usual, spherical symmetry constrains these fields to 
the following non vanishing components and coordinate dependence 
respectively
\be
Q_{ij}(\sr) & = & \diag \Big[\Lambda^2(\sr), \quad R^2(\sr), \quad 
R^2(\sr)\sin^2\st   \Big]\\
P^{ij}(\sr) & = & \sin\st \diag \Big[\frac{\PL(\sr)}{2 \Lambda(\sr)}, 
\quad \frac{\PR(\sr)}{4 R(\sr)}, \quad \frac{\PR(\sr)}{4 
R(\sr)}\sin^{-2}\st,   \Big] \quad .
\ee
Following \cite{BM87}, in particular the parity conditions derived 
there, the following decay behaviour is sufficient to guarantee a well
defined symplectic sytructure
\be \label{asymptotics}
\Lambda & \rightarrow & 1 + \frac{\mu}{\sr} + 
\mathcal{O}(\sr^{-(1+\epsilon)}) \nn \\
R & \rightarrow & \sr  + \mathcal{O}(\sr^{-(1+\epsilon)}) \nn \\
\PL & \rightarrow & \mathcal{O}(\sr^{-\epsilon}) \nn \\
\PR & \rightarrow &  \mathcal{O}(\sr^{-(1+\epsilon)}) \quad ,
\ee
In order to derive well defined equations of motion one should also 
make sure that the Hamiltonian is finite and functionally 
differentiable.
In contrast to the gauge variant framework, it is not the canonical 
Hamiltonian (with independent lapse and shift fields) but rather the 
physical Hamiltonian (with prescribed, field dependent lapse and shift)
that one has to consider. While their variations are algebraically 
almost 
identical, the field dependence of lapse and shift prevents one from 
adding the usual ADM counterterms under the usual decay behaviour. 

In \cite{ghtw1} it was already emphasised that, due to the dynamical 
nature of lapse and shift in this framework (see (\ref{lapseandshift})), 
the (geometry parts of the) diffeomorphism constraints $C_i$ must fall 
off strictly faster than the (geometry parts of the) Hamiltonian 
constraint $C$ in order to accommodate asymptotically flat spacetimes:
\be
\lim\limits_{\sr \rightarrow \infty}\frac{C_i}{C} \rightarrow 0 
\label{lim_infty}
\ee
So one has to be careful in choosing elementary fields $Q_{ij}, P^{ij}$ 
such that these conditions hold. As we have seen in the last section the 
standard conditions are generally not strong enough for this purpose, 
they only guarantee that $C\rightarrow \mathcal{O}(\sr^{-(1+\epsilon)})$ 
and $C_\sr \rightarrow \mathcal{O}(\sr^{-(1+\epsilon)})$. In this work
we did satisfy this stronger fall off behaviour by demanding 
that the shift (or equivalently the momentum density) $N^i = 
-Q^{ij}C_j/H = 0$ vanishes. It is then sufficient to add to 
the physical Hamiltonian the ADM mass term
\be
E_{ADM} & = & \frac{1}{\kappa} 8 \pi \lim\limits_{r \rightarrow 
\infty}\Big[ \Lambda^2 \sr + \frac{R^2}{\sr} - 2R R' \Big] =   \mu \quad 
.
\ee

In the general case one has to choose dynamical 
fields 
$\Lambda, R, \PL, \PR$ such that, in addition to (\ref{asymptotics}), 
also (\ref{lim_infty}) holds. This means that the physical Hamiltonian 
decays as $H \rightarrow \mathcal{O}(\sr^{-(1+\epsilon)})$ and 
asymptotically lapse and shift behave as
\be
N & := & \frac{C}{H} \rightarrow  1\\
N^i & := & -\frac{Q^{ij}C_j}{H} \rightarrow 0 \quad .
\ee

\section{Spherically Symmetric Coordinate Systems}
\label{sscs}

In the main text we have worked with various presentations 
of spherically symmetric metrics in various coordinate systems.
For the benefit of the reader we recall here how these coordinate 
systems are related with each other. Our notation is as follows:
We call $(\tau,\sr)$ the Lemaitre time and radial coordinate which 
coincide with our dust time and radial coordinate. 
Schwarzschild coordinates are denoted by $(T,R)$. The 
(generalised) Painleve -- 
Gullstrand hybrid coordinates are $(\tau,R)$. The Lemaitre, 
Schwarzschild and strict Painleve -- Gullstrand solutions are nothing 
else than coordinate transformations of the static vacuum Schwarzschild 
solution into comoving coordinates (on restricted patches of the 
fully extended Kruskal spacetime). The LTB family are not vacuum 
solutions and are expressed most easily in Lemaitre coordinates. 
Transforming the non trivial LTB solutions (i.e. non vanishing dust 
energy density) into Schwarzschild 
coordinates is not possible without picking up a non vanishing shift 
since these spacetimes are neither stationary nor static. However, there 
is a notion of semi staticity as elaborated on in the main text.

\subsection{Lema\^{i}tre solution} \label{lemaitre}
Starting from the usual Schwarzschild solution for vacuum spacetimes one 
can perform a coordinate transformation into comoving coordinates and 
arrive at what is known as the Lema\^{i}tre solution \cite{lemaitre}.

Let us start with the Schwarzschild solution in a spherical  coordinate 
chart $(T, R, \theta, \phi)$, where $T$ and $R$ approach the usual 
Minkowskian temporal and radial coordinates of an observer located at 
spatial infinity. In this coordinate system the line element can be 
written as
\be
\label{Schwds}
ds^2 & = & -\Big(1-\frac{R_s}{R}\Big)dT^2 + 
\frac{1}{1-\frac{R_s}{R}}dR^2 + R^2d\Omega^2 \quad ,
\ee
where $R_s= 2MG/c^2$ is the Schwarzschild radius, M the central mass  
and $d\Omega^2 := d\theta^2 + \sin^2\theta d\phi^2$ the area element on 
the unit sphere. That means the metric components are given by
\be
\label{Schwmetric}
g^{\rm Schw}_{\mu \nu} & := & \diag\Big[-\Big(1-\frac{R_s}{R}\Big),\quad 
\frac{1}{1-\frac{R_s}{R}},\quad R^2,\quad R^2 \sin\theta\Big] \quad ,
\ee
Then we perform the following coordinate transformation
\be
\label{diffLem}
d\tau & : = & dT + \sqrt{\frac{R_s}{R}}\Big(\frac{1}{1 - \frac{R_s}{R}} 
\Big) dR\\
d\sr & : = & dT + \sqrt{\frac{R}{R_s}}\Big(\frac{1}{1 - 
\frac{R_s}{R}}\Big)dR\quad ,
\ee
or in integrated form
\be
\label{intLem}
\tau(T,R) & = & T + 2\sqrt{R_sR} - R_s\log\left[ 
\frac{1+\sqrt{\frac{R}{R_s}}}{\big|1-\sqrt{\frac{R}{R_s}}\big|}  
\right]\\
\sr(\tau, R) & = & t(T, R) + \frac{2}{3}\sqrt{\frac{R}{R_s}}R \quad .
\ee
Obviously this coordinate transformation is only valid for $R \neq R_s$ 
where the Schwarzschild coordinates are not well defined. The inverse 
coordinate transformation is given by
\be
\label{RLemaitre}
R(\tau,\sr) & = & \Big[ 3/2\sqrt{R_s}(\sr-\tau) \Big]^{2/3}\\
T(\tau,\sr) & = & \tau - 2\sqrt{R_s R(\tau,\sr)} + \log\Big(1 + 
\sqrt{\frac{R(\tau,\sr)}{R_s}}\Big) - \log{\Big|1 - 
\sqrt{\frac{R(\tau,\sr)}{R_s}}\Big|} \quad ,
\ee
or in differential form
\be
dT & = & \frac{1}{1-\frac{R_s}{R}}d\tau + 
\frac{1}{1-\frac{R_s}{R}}d\sr\\
dR & = & -\sqrt{ \frac{R_s}{R} }d\tau + \sqrt{\frac{R_s}{R}}d\sr \quad .
\ee
This leads to the following line element for a vacuum spacetime in the 
coordinate chart $(\tau,\sr,\theta,\phi)$
\be
ds^2 & = & -d\tau^2 + \frac{R_s}{R(\tau,\sr)}d\sr^2 + 
R^2(\tau,\sr)d\Omega^2 
\quad .
\ee
This solution is known as Lema\^{i}tre metric and as one can easily read 
of the metric components in this coordinate chart 
\be
g^{Lem}_{\mu \nu} & = & \diag\Big[-1,\quad \frac{R_s}{R(\tau,\sr)}, 
\quad 
R^2(\tau,\sr), \quad R^2(\tau,\sr)\sin^2\theta\Big] \quad .
\ee
This metric describes the local coordinate system of an observer who is 
freely falling under the the influence of the central mass. It it not 
obvious (as for the Schwarzschild solution) right from the beginning 
that this solution describes a static spacetime, but one can easily 
calculate that for $R>R_s$ there exists a timelike, 
hypersurface--orthogonal Killing vector $\vec{\xi}^K \propto 
[1,1,0,0]^T$ .
{
\subsection{Painleve -- Gullstrand -- Solution}
\label{GPsol}
In order to transform the Schwarzschild solution shown in equation 
(\ref{Schwds}) and (\ref{Schwmetric}) respectively to Painleve -- 
Gullstrand coordinates, we introduce an observer that moves along 
ingoing radial, timelike geodesics of the Schwarschild spacetime. Thus 
the observer's time is identical with the Lemaitre time $\tau$ 
introduced 
in the last section in equation (\ref{diffLem}) in differential form and 
in equation (\ref{intLem}) in integrated form. In contrast to the 
Lemaitre metric the Painleve -- Gullstrand solution 
\cite{Painleve,Gullstrand} has as the radial component the Schwarzschild 
$R$. Hence, we want to perform a transformation from the Schwarzschild 
coordinates $(T,R,\theta,\phi)$ to the Painleve -- Gullstrand  
coordinates $(\tau,R,\theta,\phi)$. This coordinate transformation has 
the 
following differential form
\be
\label{diffPG}
d\tau:=dT  + \sqrt{\frac{R_s}{R}}\left(\frac{1}{1 - 
\frac{R_s}{R}}\right)dR
\ee
The vacuum spacetime in the coordinate chart $(\tau,R,\theta,\phi)$ is 
then 
given by the following line element \cite{Painleve,Gullstrand}
\be
ds^2 = -d\tau^2 + \left(dR + \sqrt{\frac{R_s}{R}}d\tau\right)^2 + 
R^2d\Omega^2
\ee
The components of the Painleve -- Gullstrand metric are given by
\be
g_{\mu\nu}^{PG} = \left(\begin{matrix} -1 & \sqrt{\frac{R_s}{R}} & 0 & 0 
\\ \sqrt{\frac{R_s}{R}} & 1 & 0 & 0 \\  0 & 0 & R^2 & 0 \\ 0 & 0 & 0& 
R^2 \sin^2\theta\end{matrix}\right)
\ee
This metric is no longer diagonal but still has a simple form. In 
particular the spatial hypersurfaces associated to this spacetime are 
flat because for $dt=0$ we get $ds^2=dR^2+R^2d\Omega^2$. All information 
about the curvature is encoded in the shift vector 
$\vec{N}=(N^R=\sqrt{R_s/R},0,0)$.
As in the case of the Lemaitre solution the coordinate transformation 
from Schwarzschild to Painleve -- Gullstrand coordinates can only be 
performed when $R\not=R_s$.
This coordinate transformation corresponds to a negative $\dot{R}$ in 
Schwarzschild coordinates (in general 
$\dot{R}=\pm\sqrt{\frac{R_s}{R}}$), $R$ decreases in time since we are 
considering observers moving along ingoing geodesics. Choosing the 
opposite sign for $\dot{R}$ yields the transformation
\be
d\tau:=dT  - \sqrt{\frac{R_s}{R}}\left(\frac{1}{1 - 
\frac{R_s}{R}}\right)dR
\ee
For this form of $dt$ we end up with the following line element
\be
ds^2 = -d\tau^2 + \left(dR - \sqrt{\frac{R_s}{R}}dt\right)^2 + 
R^2d\Omega^2
\ee
This line element corresponds to the so called "white hole region" of 
the extended Schwarzschild spacetime.
}

\subsection{Lema\^{i}tre--Tolman--Bondi solutions} \label{LTB}
The Lema\^{i}tre--Tolman--Bondi solution (LTB) is a family of exact 
solutions to Einstein's field equations (see e.g. \cite{lemaitre, 
tolman, bondi}) that describe dynamics of a spherically symmetric 
spacetime filled with inhomogeneous, pressureless dust with energy 
momentum tensor $T_{\mu \nu} = \rho_D U_\mu U_\nu$ where $\rho_D$ is the 
dust's energy density and $U_\mu = [1,0,0,0]^T$ its velocity vector 
field in comoving coordinates\footnote{See appendix \ref{covariant} for 
a covariant derivation of the equations of motion for this model.}.\\
A general solution to this problem (in a spherical coordinate chart 
$(\tau,\sr,\theta,\phi)$) is given by
\be
\label{dsLTB}
ds^2 & = & -d\tau^2 + \frac{ R^{'2}(\tau,\sr) }{ 1 + E(\sr) }d\sr^2 + 
R^2(\tau,\sr)d\Omega^2 \nn \\
\rho_D(\tau,\sr) & = & \frac{2 F'(\sr)}{R^2(\tau,\sr)R'(\tau,\sr)} \nn 
\\
\dot{R}(\tau,\sr) & = & \pm \sqrt{E(\sr) + \frac{F(\sr)}{R(\tau,\sr)}} 
\quad 
,
\ee
where slash and dot denote derivatives with respect to $\sr$ and $\tau$ 
respectively.\\
So one can characterise a particular model by choosing particular 
$\sr$-dependent functions $E(\sr)$ and $F(\sr)$. Then the physical 
radius 
$R(\tau,\sr)$ is fixed up to an arbitrary function $\beta(\sr)$ which 
characterises different initial conditions for $R(\sr,0)$. The quantity 
$E(\sr)$ determines the evolution of $R$ as well as the local geometry 
and 
$F(\sr)$ is related to the mass inside a shell with radial label $\sr$, 
see 
\cite{bondi} or for a more detailed explanation of the physical 
relevance of this model.\\
For different values of $E(\sr)$ one obtains different solutions, which 
using the notation of \cite{landau_lifshitz} can be given in parametric
form  as follows:
\begin{itemize}
\item $E(\sr) > 0$:\\
\be
R(\tau,\sr) & = & \frac{F(\sr)}{2E(\sr)}(\cosh \eta - 1) \nn \\
(\sinh \eta - \eta ) & = & \frac{2E^{3/2}(\sr)(\beta(\sr) - 
\tau)}{F(\sr)}
\ee
\item $E(\sr) = 0$:\\
\be
R(\tau,\sr) & = & \Big[ \frac{3}{2}\sqrt{F(\sr)}(\beta(\sr) - \tau)  
\Big]^{2/3} 
\ee
\item $E(\sr) < 0$: \\
\be
R(\tau,\sr) & = & \frac{F(\sr)}{2(-E(\sr))}(1 - \cos \eta) \nn \\
(\eta - \sin \eta) & = & \frac{2 (-E(\sr))^{3/2}(\beta(\sr) - 
\tau)}{F(\sr)}
\ee
\end{itemize}
The LTB--family of solutions is widely used to describe astrophysical 
situations, for example it can be used to model the gravitational 
collapse of a (only gravitationally interacting) matter cloud. 
Furthermore it can be applied to cosmological models which go beyond the 
standard assumptions of homogeneity in ordinary FRW--evolution.
{
\subsection{Generalised Painleve -- Gullstrand Solutions}
\label{GPGSol}
Following \cite{LaskyLunBurston} we present in this section how the LTB 
metric can be transformed into a metric expressed in terms of 
generalised Painleve -- Gullstrand coordinates (GPG). As for the 
ordinary Painleve -- Gullstrand coordinates we want the LTB -- time and  
the GPG -- time to coincide and for the radial component we take the 
function $R(\tau,\sr)$ occurring in front of $d\Omega^2$ in the LTB line 
element in equation (\ref{dsLTB}). Hence, the transformation from LTB -- 
coordinates $(\tau,\sr,\theta,\phi)$ to GPG -- coordinates denoted by 
$(\tau,R,\theta,\phi)$ is in differential form given by
\be
\label{dRGPG}
dR = \frac{\partial R}{\partial \tau}d\tau + \frac{\partial R}{\partial 
\sr}d\sr 
= 
\dot{R}d\tau + R^{\prime}d\sr
\ee
Let us consider the following general ansatz for the line element
\be
ds^2 = -Xd\tau^2 + YdR^2 +Zd\tau dR +R^2d\Omega^2 
\ee
where $X,Y,Z$ are functions of $\tau$ and $R$. Considering the explicit 
form of $dR$ in equation (\ref{dRGPG}) and comparing with the LTB line 
element in equation (\ref{dsLTB})
we obtain the following conditions for the functions $X,Y$ and $Z$
\begin{eqnarray}
{
X-Y\dot{R}^2-Z\dot{R} } &=& 1 \\
YR^{\prime 2} &=& \frac{R^{\prime 2}}{1+E} \\
ZR^{\prime}+ 2Y\dot{R}R^{\prime} &=& 0
\end{eqnarray}
This system of equations has the following solutions for $X,Y$ and $Z$
\be
{
X(\tau,R)=1-\frac{\dot{R}^2}{1+E(\tau,R)}   },\quad 
Y(t,R)=\frac{1}{1+E(\tau,R)},\quad 
Z(\tau,R)=-\frac{2\dot{R}}{1+E(\tau,R)}
\ee
where substitution for $\dot{R}$ via (D.1) is being understood.
The function $E(\sigma_r)$ occurring in the LTB solution becomes a 
function of $E(\tau,R)$ when expressing $\sr$ in terms of $\tau,R$.
The equation for 
$\dot{R}^2$ in LTB coordinates given in equation (\ref{dsLTB}) has the 
following expression in the GPG coordinates
\be
\dot{R}^2 = E(\tau,R) + \frac{F(\tau,R)}{R}\quad \Leftrightarrow \quad 
\dot{R}=\pm\sqrt{E(\tau,R) + \frac{F(\tau,R)}{R}}.
\ee
Since we want to cover the black hole region of the extended 
Schwarschild spacetime, we choose similar to the case of the ordinary GP 
coordinates a negative square root for $\dot{R}$ and obtain the 
following line element
\be
\label{AppdsGPG}
ds^2 = -d\tau^2 +\frac{\left(dR +\sqrt{E(\tau,R) + 
\frac{F(\tau,R)}{R}}d\tau\right)^2}{1+E(\tau,R)} + R^{2}d\Omega^2
\ee
This line element is well defined for all values of $\tau,R$ for which 
$E(\tau,R)>-1$ and $E(\tau,R) + F(\tau,R)/R\ge 0$.
The opposite sign choice for $\dot{R}$ leads to a minus sign in front of 
the $dt$ -- term in the bracket of the second term and describes, as 
before, the generalised "white hole region" of the spacetime. 
\\
Now when solving the ADM equations that lead to the GPG metric one 
obtains an evolution equation for the shift vector 
$\vec{N}(\tau,R)=(N^R(\tau,R),0,0)$ 
given by 
\be
{\cal L}_{\vec{n}}\left((N^{R}(\tau,R))^2 - \frac{F(\tau,R)}{R}\right)=0 
\quad\mathrm{with}\quad n^{\mu}=(1,-N^R,0,0)
\ee
being the unit vector normal to the spatial hypersurfaces of that 
spacetime. 
This is a second order equation for $N^R$ because as shown in 
\cite{LaskyLunBurston} the term $F(\tau,R)/R$ is related to the shift 
vector component by $2R{\cal L}_{\vec{n}}(N^R)(\tau,R)=F(\tau,R)/R$.
See \cite{LaskyLunBurston} for more details. From equation 
(\ref{AppdsGPG}) we can easily read off the explicit form of the shift 
vector
\be
\label{NRGPG}
N^R(\tau,R) = \sqrt{E(\tau,R) + \frac{F(\tau,R)}{R}}
\ee
Consequently, the equation for the lapse carries over to an equation for 
$E(\tau,R)$ given by
\be
\label{PartE}
{\cal L}_{\vec{n}}\left(E\right)(\tau,R)=0\quad\Leftrightarrow\quad 
\frac{\partial E(\tau,R)}{\partial \tau} -\sqrt{E(\tau,R) + 
\frac{F(\tau,R)}{R}}\frac{\partial E(\tau,R)}{\partial R} =0
\ee
Furthermore, by applying the Lie derivative ${\cal L}_{\vec{n}}$ onto 
$(N^R)^2$ and using its explicit expression in terms of $E,F$ and $R$ 
shown in equation (\ref{NRGPG}) we obtain
\be
{\cal L}_{\vec{n}}(F) = N^R\left(2R{\cal L}_{\vec{n}}(N^R)(\tau,R) - 
\frac{F(\tau,R)}{R}\right)=0
\ee
which yields a partial differential equation for $F(\tau,R)$
\be
\label{PartF}
 \frac{\partial F(\tau,R)}{\partial \tau} -\sqrt{E(\tau,R) + 
\frac{F(\tau,R)}{R}}\frac{\partial F(\tau,R)}{\partial R} =0.
\ee
Consequently, for the GPG metric only those functions $E$ and $F$ are 
allowed that satisfy the partial differential equations in (\ref{PartE}) 
and (\ref{PartF}).

\section{Covariant analysis of spherically symmetric gravity plus 
pressureless dust} \label{covariant}

For the sake of completeness,
we analyse the dynamics of pressureless dust--matter 
coupled to gravity in a spherically symmetric setting using the usual 
covariant framework of GR. The metric $g_{\mu \nu}$ is a priori 
not a gauge invariant object and the ``dynamics'' generated by 
Einstein's equations has to be interpreted as gauge transformations in 
the strict sense. In order to make the framework physically meaningful,
we must fix the spacetime diffeomorphism freedom. We therefore fix a 
comoving 
coordinate system with respect to which the lapse is unity and the shift 
vanishes. In the Hamiltonian language, this completely fixes the 
gauge freedom generated 
by the radial diffeomorphism and Hamiltonian constraints respectively.
The comoving coordinates will be denoted by $(t,r)$ in order to 
emphasise that they are measuring proper time $t$ along {\it ideal test
observer} geodesics labelled by $r$. In the usual LTB framework, and in 
contrast to the Brown -- Kucha\v{r} framework, there 
is no ``observer Lagrangian'' that actually models these observers 
and their graviational interaction and whose proper time and geodesic 
label we have denoted $(\tau,\sr)$ throughout the text. Rather, one 
constructs a spherically symmetric, pressure free, perfect fluid 
energy momentum tensor whose Lagrangian origin remains obscure and whose 
dynamics is simply induced by the Bianchi identity of the Einstein
equations. 

In what follows we will see that one ends up with 
an exact mathematical match between the two frameworks although their
conceptual starting points are quite different, upon identifying 
$(\tau,\sr):=(t,r)$. At first sight this may seem mathematically not too 
surprising 
because the energy momentum tensor of the Brown -- Kucha\v{r} Lagrangian 
has a perfect fluid form whose pressure is constrained\footnote{That is,
the vanishing of the pressure is an equation of motion and not put in by 
hand.} 
to vanish.  
However, since the Brown -- Kucha\v{r} Lagrangian involves 
altogether eight dust fields to begin with and displays a complicated 
gauge symmetry involving first and second class constraints, it is 
after all not straightforward to see that one obtains a perfect match.
In particular, the velocity field of the Brown -- Kucha\v{r} Lagrangian
is a complicated aggregate composed out of dust fields and a 
priori cannot be 
prescribed to take a distinguished form.\\
\\
We will assume spherical symmetry, therefore we can make the following 
ansatz for the line--element $ds^2$:
\be
ds^2 & = & -dt^2 + \Lambda^2(t,r)dr^2 + R^2(t,r)d\theta^2 
+R^2(t,r)\sin^2\theta d\phi^2 \quad .
\ee
This means the metric components are given by:
\be
g_{\mu \nu} & = & \diag\Big[ -1,\quad \Lambda^2(t,r), \quad R^2(t,r), 
\quad R^2(t,r)\sin^2\theta    \Big] \quad
\ee
Now we compute the Christoffel--symbols $\Gamma^{\lambda}_{\mu \nu } := 
\frac{1}{2}g^{\lambda \sigma} \Big[ g_{\sigma \mu, \nu} + g_{\sigma \nu, 
\mu} - g_{\mu \nu, \sigma}   \Big]$ of $g_{\mu \nu}$ where we used the 
abbreviation $g_{\mu \nu, \sigma} := \frac{\partial}{\partial 
x^{\sigma}}g_{\mu \nu}$.\\
The Christoffel--symbols are given by:
\be
\Gamma^{t} & = &
\begin{pmatrix}
0 & 0 & 0 & 0\\
0 & \Lambda \dot{\Lambda} & 0 & 0\\
0 & 0 & R \dot{R} & 0\\
0 & 0 & 0 & R \dot{R}\sin^2\theta 
\end{pmatrix}\\
\;\\
\Gamma^{r} & = &
\begin{pmatrix}
0 & \frac{\dot{\Lambda}}{\Lambda} & 0 & 0\\
\frac{\dot{\Lambda}}{\Lambda} & \frac{\Lambda'}{\Lambda} & 0 & 0\\
0 & 0 & -\frac{R R'}{\Lambda^2} & 0 \\
0 & 0 & 0 & -\frac{R R'}{\Lambda^2}\sin^2\theta
\end{pmatrix}\\
\;\\
\Gamma^{\theta} & = &
\begin{pmatrix}
0 & 0 & \frac{\dot{R}}{R} & 0\\
0 & 0 & \frac{R'}{R} & 0\\
\frac{\dot{R}}{R} & \frac{R'}{R} & 0 & 0\\
0 & 0 & 0 & -\sin\theta\cos\theta
\end{pmatrix}\\
\;\\
\Gamma^{\phi} & = &
\begin{pmatrix}
0 & 0 & 0 & \frac{\dot{R}}{R}\\
0 & 0 & 0 & \frac{R'}{R}\\
0 & 0 & 0 & \cot\theta\\
\frac{\dot{R}}{R} & \frac{R'}{R} & \cot\theta & 0
\end{pmatrix} \quad .
\ee
This leads to the following non--vanishing components of the 
Ricci--curvature tensor\\ $R_{\mu \nu} = \Gamma_{\mu \nu, \rho}^{\rho} - 
\Gamma_{\mu \rho, \nu}^{\rho}   + \Gamma_{\mu \nu}^\rho \Gamma_{\rho 
\sigma}^{\sigma} - \Gamma_{\mu \rho}^{\sigma} \Gamma_{\nu 
\sigma}^{\rho}$:
\be
R_{tt} & = & - \frac{\ddot{\Lambda}}{\Lambda}  -  2\frac{\ddot{R}}{R}  
\nn \\
R_{rr} & = & \Lambda \ddot{\Lambda}  -  2 \frac{R''}{R}  + \frac{2 
\Lambda \dot{\Lambda} \dot{R}}{R}  +  \frac{2 \Lambda' R'}{\Lambda R} 
\nn  \\
R_{\theta\theta} & = & R \ddot{R}  -  \frac{R R''}{\Lambda^2}  +  
\dot{R}^2  -  \frac{R^{'2}}{\Lambda^2}  +  \frac{R \Lambda' 
R'}{\Lambda^3}  +  \frac{R \dot{\Lambda} \dot{R}}{\Lambda}  +  1 \nn \\
R_{\phi\phi} & = & \sin^2\theta R_{\theta\theta} \nn \\
R_{tr} & = & -2 \frac{\dot{R}'}{R}  +  2 \frac{\dot{\Lambda} R' 
}{\Lambda R} \quad .
\ee
The Ricci--scalar $R := g^{\mu \nu} R_{\mu \nu}$ then takes the form
\be
R & = & 2 \frac{\ddot{\Lambda}}{\Lambda}  +  4 \frac{\ddot{R}}{R}  -4 
\frac{R''}{\Lambda^2 R}  + 2 \Big( \frac{\dot{R}}{R}   \Big)^2  -  
\frac{2}{\Lambda^2} \Big( \frac{R'}{R}  \Big)^2  + 4 \frac{\dot{\Lambda} 
\dot{R}}{\Lambda R}  +  4 \frac{\Lambda' R'}{\Lambda^3 R}  + 
\frac{2}{R^2}
\ee
Now we want to couple matter to the system, more precisely we will use 
inhomogeneous pressureless dust--matter. Assuming spherical symmetry the 
pressure free dust stress--energy tensor is given by
\be
T_{\mu \nu} & = & \rho_{\rm D}(t,r) U_\mu U_\nu \quad ,
\ee
where $\rho_{\rm D}(t,r)$ can be interpreted as the dust energy 
density and $U^\mu$ is the dust velocity vector field which in comoving 
coordinates takes the form $U^\mu = \delta_t^\mu$.\\
Having collected all ingredients we can now write down Einstein's 
equations $R_{\mu \nu} - 1/2 \,^{(4)}R g_{\mu \nu} = \kappa/2 T_{\mu 
\nu}$ with $\kappa = 16 \pi G /c^4$ for the system under consideration:
\be
-\frac{2 R''}{\Lambda^2 R}  -  \frac{1}{\Lambda^2}\Big( \frac{R'}{R}  
\Big)^2  +  \Big( \frac{\dot{R}}{R} \Big)^2  +  2 \frac{\Lambda' 
R'}{\Lambda^3 R}  +  2 \frac{\dot{\Lambda} \dot{R}}{\Lambda R}  +  
\frac{1}{R^2}  & = & \kappa \frac{\rho_D}{2}    \qquad \mbox{\small 
$tt$-comp.} \label{tt}  \nn \\
-2 \frac{\ddot{R} \Lambda^2 }{R}  -  \Lambda^2 \Big( \frac{\dot{R}}{R}  
\Big)^2  +  \Big( \frac{R'}{R}  \Big)^2  -  \frac{\Lambda^2}{R^2}  & = & 
0        \qquad \mbox{\small $rr$-comp.} \label{rr} \nn \\
-R \ddot{R}  +  \frac{R R''}{\Lambda^2}  -  \frac{R^2 
\ddot{\Lambda}}{\Lambda}  -  \frac{R \Lambda'  R'}{\Lambda^3}  -  
\frac{R \dot{\Lambda} \dot{R}}{\Lambda}    & = & 0    \qquad 
\mbox{\small $\theta\theta$- and $\phi\phi$-comp.} \label{thetatheta} 
\nn \\
\dot{R}'  -  \frac{\dot{\Lambda} R'}{\Lambda}  & = & 0 \qquad 
\mbox{\small $tr$-comp.}\label{tr} \quad .
\ee
Imposing the Bianchi identity (energy momentum conservation) 
we find 
\be 
\nabla^\mu T_{\mu\nu}=-\delta^t \rho_{{\rm D}}
\frac{\partial}{\partial t}[\ln(\rho_{{\rm D}} \Lambda R^2)]=0
\ee
which obviously constrains $\rho_D$ to have 
the form $-\epsilon(r)/[\Lambda(t,r) R^2(t,r)]$ for some free 
function $\epsilon$ of $r$ only.

One can easily see that this system of partial differential equations 
coincides formally with the one obtained in section \ref{eoms}, 
equations (\ref{2ndI}) - (\ref{2ndIV}). So we can map the problem of 
finding spherically symmetric solutions in the framework of gauge 
invariant observables to the problem of finding spherically symmetric 
solutions for gravity coupled to pressureless dust--matter in the usual 
framework of Einstein's General Relativity in the comoving gauge
$N=1,N^r=0$.\\
The system of solutions to this system of partial differential equations 
is the so called Lema\^{i}tre-Tolman-Bondi family (see appendix 
\ref{LTB}).
\bibliographystyle{utphys}
\bibliography{pa84pub}

\end{document}